\documentclass[aps,preprint,floats,epsf,epsfig,nofootinbib]{revtex4}
\usepackage{epsfig}
\usepackage{graphicx}

\begin{document}
\baselineskip=20 pt
\def\l{\lambda}
\def\m{\mu}
\def\L{\Lambda}
\def\bt{\beta}
\def\mphi{m_{\phi}}
\def\hphi{\hat{\phi}}
\def\vphi{\langle \phi \rangle}
\def\etamunu{\eta^{\mu\nu}}
\def\dmul{\partial_{\mu}}
\def\dnul{\partial_{\nu}}
\def\bea{\begin{eqnarray}}
\def\eea{\end{eqnarray}}
\def\bfl{\begin{flushleft}}
\def\efl{\end{flushleft}}
\def\bc{\begin{center}}
\def\ec{\end{center}}
\def\bcr{\begin{center}}
\def\ecr{\end{center}}
\def\al{\alpha}
\def\bt{\beta}
\def\eps{\epsilon}
\def\lam{\lambda}
\def\gam{\gamma}
\def\s{\sigma}
\def\r{\rho}
\def\e{\eta}
\def\dl{\delta}
\def\non{\nonumber}
\def\nont{\noindent}
\def\la{\langle}
\def\ra{\rangle}
\def\nc{{N_c^{\rm eff}}}
\def\vp{\varepsilon}
\def\drho{\bar\rho}
\def\deta{\bar\eta}
\def\vma{{_{V-A}}}
\def\vpa{{_{V+A}}}
\def\J{{J/\psi}}
\def\ov{\overline}
\def\Lqcd{{\Lambda_{\rm QCD}}}
\def\btr{\bigtraingleup}
\def\pr{{\sl Phys. Rev.}~}
\def\prl{{\sl Phys. Rev. Lett.}~}
\def\pl{{\sl Phys. Lett.}~}
\def\np{{\sl Nucl. Phys.}~}
\def\zp{{\sl Z. Phys.}~}


\title{Data for Polarization in Charmless $B \to \phi
K^{*}$: A Signal for New Physics?}


\author{\large\sl Prasanta~Kumar~Das and Kwei-Chou~Yang}

\address{Department of Physics, Chung-Yuan Christian University, Chung-Li,
Taiwan 320, R.O.C.}

 \vskip1cm \small
\begin{abstract}  The recent observations of sizable
transverse fractions of $\overline B\to\phi \overline K^*$ may
hint for the existence of new physics. We analyze all possible
new-physics four-quark operators and find that two classes of
new-physics operators could offer resolutions to the $\overline
B\to \phi \overline K^*$ polarization anomaly. The operators in
the first class have structures $(1-\gamma_5)\otimes (1-\gamma_5),
\sigma(1-\gamma_5)\otimes \sigma(1-\gamma_5)$, and in the second
class $(1+\gamma_5)\otimes (1+\gamma_5), \sigma(1+\gamma_5)\otimes
\sigma(1+\gamma_5)$. For each class, the new physics effects can
be lumped into a single parameter. Two possible experimental
results of polarization phases,
$\arg(A_\perp)-\arg(A_\parallel)\approx\pi$ or $0$, originating
from the phase ambiguity in data, could be separately accounted
for by our two new-physics scenarios: the first (second) scenario
with the first (second) class new-physics operators.  The
consistency between the data and our new physics analysis,
suggests a small new-physics weak phase, together with a large(r)
strong phase. We obtain sizable transverse fractions
$\Lambda_{\parallel\parallel} + \Lambda_{\perp\perp}\approx
\Lambda_{00}$, in accordance with the observations. We find
$\Lambda_{\parallel\parallel} \simeq 0.8 \Lambda_{\perp\perp}$ in
the first scenario but $\Lambda_{\parallel\parallel} \gtrsim
\Lambda_{\perp\perp}$ in the second scenario. We discuss the
impact of the new-physics weak phase on observations.
\end{abstract}

\maketitle

\section{Introduction}
The studies for two-body charmless $B$ decays  have raised a lot
of interests among the particle physics community. Recently the
BABAR and BELLE collaborations have presented important results
for the $B$ meson decaying to a pair of light vector mesons (with
$V = \phi,\rho$, or
$K^*$)~\cite{Aubert:2003mm,Aubert:2003xc,Aubert:2004xc,Zhang:2003up,Chen:2003jf,Abe:2004ku}.
This immediately surges a considerable amount of theoretical
attentions to study nonperturbative features or to look for the
possibility of having new physics~(NP) in order to explain several
discrepancies between the data and the Standard Model~(SM) based
calculations~\cite{Kagan,Hnli,Colangelo:2004rd,Cheng:2004ru,Ladisa:2004bp,Dariescu:2003tx,Hou:2004vj}.

From the existing SM calculations for the charmless $\overline B
\rightarrow VV$ modes, it is known that the amplitude, $\overline
H_{00}$ ($\sim {\cal O}(1)$), with two vector mesons in the
longitudinal polarization state is much greater than those in
transverse polarization states,  since the latter are found to be
$\overline H_{--}\sim {\cal O}(1/m_b), \overline H_{++}\sim{\cal
O}(1/m_b^2)$ (or $\overline A_{\|}\simeq \overline A_{\perp}\sim
{\cal O}(1/m_b)$ in the transversity basis)~\cite{CY}. For the $B$
meson decays, the relation for different helicity amplitudes is
modified as $H_{00}: H_{++}: H_{--}\sim {\cal O}(1): {\cal
O}(1/m_b): {\cal O}(1/m_b^2)$. Nevertheless, recently
BABAR~\cite{Aubert:2003mm,Aubert:2004xc} first observed sizable
transverse fractions in the $\overline B \to \phi \overline K^*$
decays, where the transverse polarization amplitudes are
comparable to the longitudinal one. This result was confirmed
later by BELLE~\cite{Chen:2003jf,Abe:2004ku}. In other words, in
terms of helicity amplitudes the data show that $|\overline H_{++}
\pm \overline H_{--}|^2 \approx |\overline H_{00}|^2 $ (or
$2|\overline A_{\|}|^2\approx 2|\overline A_{\perp}|^2\approx
|\overline A_0|^2$ in the transversity basis). Such an anomaly in
transverse fractions is rather unexpected within the SM framework.
Efforts have already been made for finding a possible explanation
in the SM or NP scenario. In the SM, according to
Kagan~\cite{Kagan}, the nonfactorizable contributions due to the
annihilation could give rise to the following logarithmic
divergent contributions to the helicity amplitudes: ${\overline
H}_{00}, {\overline H}_{--} \sim {\cal O}[(1/m_b^2) \ln^2
(m_b/\Lambda_{h})]$, ${\overline H}_{++} \sim {\cal O}[(1/m_b^4)
\ln^2 (m_b/\Lambda_{h})]$, where $\Lambda_{h}$ is the typical
hadronic scale. This in turn may enhance the transverse amplitudes
required to explain the anomaly. However, in the perturbative QCD
(PQCD) framework, Li and Mishima~\cite{Hnli} have shown that the
annihilations are still not sufficient to enhance transverse
fractions. Another possibility for explaining the polarization
anomaly advocated by Colangelo {\it et
al.}~\cite{Colangelo:2004rd} is the existence of large charming
penguin and final state interaction (FSI) effects. However they
got $|A_0|^2(B\to \rho K^*)< |A_0|^2 (B\to \phi K^*)$, in contrast
to the observations
\cite{Aubert:2003mm,Aubert:2003xc,Aubert:2004xc,Zhang:2003up,Chen:2003jf,Abe:2004ku},
where the normalization $\sum_i |A_i|^2=1 $ is adopted. With the
similar FSI scenario, Cheng {\it et al.}~\cite{Cheng:2004ru}
obtained $|A_0|^2: |A_{\|}|^2: |A_\perp|^2=0.43:0.54:0.03$, which
is also in contrast to the recent
data~\cite{Aubert:2004xc,Abe:2004ku}. Now the question is: Is it
possible to explain this anomaly by the NP? If yes, what types of
NP operators one should consider? Some NP related models have been
proposed~\cite{Dariescu:2003tx}, where the so-called right-handed
currents $\bar s\gamma_{\mu}(1+\gamma_5)b\ \bar s\gamma^{\mu}(1\pm
\gamma_5)s$ were emphasized~\cite{Kagan}. If the right-handed
currents contribute constructively to $\overline A_\perp$ but
destructively to $\overline A_{0,\|}$, then one may have larger
$|\overline A_\perp / \overline A_0 |^2$ to account for the data.
However the resulting $|\overline A_{\|}|^2 \ll |\overline
A_\perp|^2$~\cite{Kagan} will be in contrast to the recent
observations~\cite{Aubert:2004xc,Abe:2004ku}. See also the
detailed discussions in Sec.~\ref{sec:section2}.

In the present study, we consider general cases of 4-quark operators.
Taking into account all possible color and Lorentz
structures, totally there are 20 NP four-quark operators which do
not appear in the SM effective Hamiltonian (see
Eqs.~(\ref{eq:vectorop}) and (\ref{eq:scalarop})). After analyzing
the helicity properties of quarks arising from various four-quark
operators, we find that only two classes of four-quark operators
are relevant in resolving the transverse anomaly. The first class
is made of operators with structures $\sigma(1-\gamma_5)\otimes
\sigma(1-\gamma_5)$ and $(1-\gamma_5)\otimes (1-\gamma_5)$, which
contribute to different helicity amplitudes as ${\overline
H_{00}}:{\overline H_{--}}:{\overline H_{++}}
  \sim  {\cal O}(1/m_b):{\cal O}(1/m_b^2):{\cal
O}(1)$. The second class consists of operators with structures
$\sigma(1+\gamma_5)\otimes \sigma(1+\gamma_5)$ and
$(1+\gamma_5)\otimes (1+\gamma_5)$, from which the resulting
amplitudes read as ${\overline H_{00}}:{\overline
H_{++}}:{\overline H_{--}} \sim {\cal O}(1/m_b):{\cal
O}(1/m_b^2):{\cal O}(1)$. Moreover, the above (pseudo-)scalar
operators can be written in terms of their companions, the
(axial-)tensor operators, by Fierz transformation. Finally, there
is only one effective coefficient relevant for each class. We find
that these two classes can separately satisfy the two possible
solutions for polarization phase data, which is due to the phase
ambiguity in the measurement, and the anomaly for large transverse
fractions can thus be resolved. The tensor operator effects were
first noticed by Kagan~\cite{Kagan} (see Sec.~\ref{sec:section2}
for further discussions).

 The organization of the paper is as follows. In Sec.~\ref{sec:section2},
we first introduce the SM results for the polarization amplitudes
in the ${\overline B^0} \to \phi {\overline K}^{*0}$ decay within
the QCD factorization (QCDF) framework. After that we give a
detailed discussion about how the NP can play a crucial role in
resolving the large transverse polarization anomaly as observed by
BELLE and BABAR. The reason for choosing the two classes of
operators with structures (i) $\sigma(1-\gamma_5)\otimes
\sigma(1-\gamma_5), (1-\gamma_5)\otimes (1-\gamma_5)$ and (ii)
$\sigma(1+\gamma_5)\otimes \sigma(1+\gamma_5), (1+\gamma_5)\otimes
(1+\gamma_5)$ is explained and the relevant calculations arising
from these operators are performed. We discuss the possibility for
the existence of right-handed currents $\bar s \gamma_\mu
(1+\gamma_5)b\ \bar s \gamma^\mu (1\pm\gamma_5)s $ which was
emphasized in~\cite{Kagan}. From the point of view of helicity
conservation in the strong interactions, we discuss various
contributions originating from the chromomagnetic dipole operator,
charming penguin mechanism, and annihilations. Some observables
relevant in our numerical analysis are defined in this section. In
Sec.~\ref{sec:section3}, we summarize input parameters {\it e.g.}
Kobayashi-Maskawa (KM) elements, form factors, meson decay
constants, required for our study.
Sec.~\ref{sec:section4} is fully devoted to the numerical analysis.
We discuss in detail two scenarios, which are separately
consistent with the two possible polarization phase solutions in
data due to the phase ambiguity. We obtain the best fit values for
the NP parameters which can resolve the polarization anomaly.
Numerical results for observables are collected in this section.
Finally, in Sec.~\ref{sec:section5}, we summarize our results and
make our conclusion.

\section{Framework} \label{sec:section2}
\subsection{The Standard Model results in the QCD factorization approach}
The best starting point for describing nonleptonic charmless $B$
decays is to write down first the effective Hamiltonian describing
the processes. The processes of our concern are the $\overline B \to \phi
\overline K^*$ decays which are penguin dominated. In the SM, the relevant
effective weak Hamiltonian  ${\cal H}_{\rm eff}$ for the above
$\Delta B = 1$ transitions is
 \bea
\label{eff}
 {\cal H}_{\rm eff} = \frac{G_F}{\sqrt{2}} \left[V_{ub} V_{us}^*
\left(c_1 O_1^u + c_2 O_2^u \right) + V_{cb} V_{cs}^* \left(c_1
O_1^c + c_2 O_2^c \right) - V_{tb} V_{ts}^* \left( \sum_{i =
3}^{10} c_i O_i \right) + c_g O_g\right] + {\rm H.c.}. \eea
 Here $c_i$'s are the Wilson coefficients and the 4-quarks
current-current, penguin and chromomagnetic dipole operators are
defined by
 \begin{itemize}
\item {\bf current-current operators}:
 \bea O_1^u &=&
({\overline{u}} b)_{V-A} ({\overline {s}} u)_{V-A}
 ~~ ~~ O_2^u = ({\overline{u}_{\al}} b_{\bt})_{V-A} ({\overline {s}}_\bt
u_\al)_{V-A}, \nonumber\\
O_1^c &=& ({\overline{c}} b)_{V-A} ({\overline {s}} c)_{V-A}
 ~~ ~~ O_2^c = ({\overline{c}_{\al}} b_{\bt})_{V-A} ({\overline {s}}_\bt
c_\al)_{V-A},
 \eea
 \item {\bf QCD-penguin operators}:
  \bea O_3 &=& ({\overline{s}} b)_{V-A}
\sum_q({\overline {q}} q)_{V-A}, ~~ ~~ O_4 = ({\overline{s}_{\al}}
b_{\bt})_{V-A} \sum_q({\overline {q}}_\bt q_\al)_{V-A},
 \nonumber\\
  O_5 &=& ({\overline{s}} b)_{V-A}
\sum_q({\overline {q}} q)_{V+A}, ~~ ~~ O_6 = ({\overline{s}_{\al}}
b_{\bt})_{V-A} \sum_q({\overline {q}}_\bt q_\al)_{V+A}, \eea
 \item {\bf electroweak-penguin operators}:
 \bea
 O_7 &=& \frac{3}{2}({\overline{s}} b)_{V-A} \sum_q e_q({\overline {q}}
q)_{V+A}, ~~ ~~ O_8 = \frac{3}{2}({\overline{s}_{\al}}
b_{\bt})_{V-A} \sum_q e_q ({\overline {q}}_\bt q_\al)_{V+A},
 \nonumber\\
  O_9 &=& \frac{3}{2}({\overline{s}} b)_{V-A}
\sum_q e_q ({\overline {q}} q)_{V-A}, ~~ ~~ O_{10} =
\frac{3}{2}({\overline{s}_{\al}} b_{\bt})_{V-A} \sum_q e_q
({\overline {q}}_\bt q_\al)_{V-A},
\eea
 \item {\bf chromomagnetic dipole operator}:
\bea O_{8 g}=\frac{g_s}{8 \pi^2} m_b {\overline s} \sigma^{\mu\nu}
(1+\gamma_5) T^a b G^a_{\mu\nu}, \eea
\end{itemize}
where $\al, \bt$ are the $SU(3)$ color indices, $V\pm A$
correspond to $\gamma^\mu (1 \pm \gamma^5)$, the Wilson
coefficients $c_i$'s are evaluated at the scale $\mu$, $e$ and $g$
are respectively QED and QCD coupling constants and $T^a$'s are
$SU(3)$ color matrices. For the penguin operators, $O_3,
\dots,O_{10}$, the sum over $q$ runs over different quark flavors,
active at $\mu \simeq m_b$, i.e. $ q~ \epsilon \{u, d, s, c, b\}$.

In the present work, we will embark on the study of
$\overline B \to \phi\overline K^*$ decays in the approach of the QCDF.
The ${\overline B}^0 \rightarrow \phi {\overline K^{*0}}$
decay amplitude with the $\phi$ meson being factorized~\cite{CY} reads
 \bea \label{Amp}
{\overline A}({\overline B}^0 \rightarrow \phi {\overline
K^{*0}})_{SM} =\frac{G_F}{\sqrt 2} \left(- V_{tb} V_{ts}^*\right)
\left[a_3 + a_4 + a_5 - \frac{1}{2}(a_7 + a_9 + a_{10})\right]
X^{({\overline B}^0 {\overline K^{*0}}, \phi)},
 \eea
 which is penguin dominated. The annihilation contribution which
is power suppressed is neglected here~\cite{CY}.
The $B^0 \rightarrow \phi K^{*0}$ decay amplitude can be obtained by
considering CP transformation. As far as the charged $B$-meson
decay is concerned, the dominant contribution also comes from the
penguin operators, while the contribution due to $O_1,O_2$ is
color and KM suppressed. In the scenario, where $\phi$ is
factorized, the decay amplitudes for $B^0, {\overline B^0}, B^+,
B^-$  are almost the same. The factor $X^{({\overline B}^0
{\overline K^{*0}}, \phi)}$ in Eq.~(\ref{Amp}) is equal to
 \bea
 X^{({\overline B}^0 {\overline K^{*0}}, \phi)}
 &=& \langle\phi(q,\eps_1)| ({\overline{s}}s)_{V-A}|0\rangle
 \langle{\overline K^{*0}} (p^\prime, \eps_2)|({\overline{s}}b)_{V-A}|
 {\overline B}^0 (p) \rangle,
\nonumber \\
&=& i f_\phi m_\phi  \left[\frac{- 2 i}{m_B + m_{K^*}}
\eps_{\mu\nu\al\bt} \eps_1^{*\mu} \eps_2^{*\nu} p^\al p^{\prime
\bt} V(q^2) \right]
\nonumber \\
&& - i f_\phi m_\phi \left[(m_B + m_{K^*}) \eps_1^*\cdot \eps_2^*
A_1(q^2) - (\eps_1^*\cdot p) (\eps_2^*\cdot p) \frac{2
A_2(q^2)}{m_B + m_{K^*}} \right], \eea
 where the decay constants and form factors are defined by
\bea
 \langle\phi(q,\eps_1)|V^\mu|0\rangle &=& f_\phi m_\phi
\eps_1^{\mu *} ,
\nonumber \\
\langle{\overline K^{*0}}(p^\prime, \eps_2)|V^\mu|{\overline B}^0 (p)\rangle
 &=& \frac{2 }{m_B + m_{K^*}} \eps^{\mu\nu\al\bt} \eps_{2 \nu}^{*} p_\al
p_{\bt}^{\prime} V(q^2),
\nonumber \\
\langle{\overline K^{*0}}(p^\prime, \eps_2)|A^\mu|{\overline B}^0 (p)\rangle
 &=& i \left[(m_B + m_{K^*}) \eps_2^{*\mu} A_1(q^2) - (\eps_2^* \cdot p)
(p + p^\prime)^\mu
\frac{ A_2(q^2)}{m_B + m_{K^*}}\right]
\nonumber \\
&& - 2 i m_{K^*} \frac{\eps_2^*\cdot p}{q^2} q^\mu \left[A_3(q^2)
- A_0(q^2)\right],
 \eea
 with $m_B$ and $m_{K^*}$ being the masses of ${\overline B^0}$
and ${\overline K^{*0}}$
mesons, respectively, $q = p - p^\prime$, $A_3(0) = A_0(0)$, and
\bea
 A_3(q^2)= \frac{m_B + m_{K^*}}{2 m_{K^*}} A_1(q^2) - \frac{m_B -
m_{K^*}}{2 m_{K^*}} A_2(q^2).
 \eea
 It is straightforward to write down the
decay width,
 \bea
  \Gamma (\overline B^0 \rightarrow \phi
{\overline K^{*0}}) = \frac{p_c}{8 \pi m_B^2}
 \left( |{\overline H_{00}}|^2 + |{\overline H_{++}}|^2
 + |{\overline H_{--}}|^2\right),
 \eea
 where $p_c$ is the center mass momentum of the
$\phi$ or $\overline K^{*0}$ meson in the $\overline B$ rest
frame. ${\overline H_{00}}$, ${\overline H_{++}}$, ${\overline
H_{--}}$ are the decay amplitudes in the helicity basis and in
QCDF~\footnote{\label{foot:1} We choose the coordinate systems in
the Jackson convention, consistent with what BaBar and Belle
did~\cite{convention}. In the $\overline B$ rest frame, if the $z$
axis of the coordinate system is along the the direction of the
flight of the $\phi$ meson and the transverse polarization vectors
of $\phi$ are chosen to be $\epsilon_{\phi}^{\mu}(\pm 1)=(0, \mp
1, -i, 0)/\sqrt{2}$, then the transverse polarization vectors of
$K^*$ are given by $\epsilon_{K^*}^{\mu}(\pm 1)=(0, \mp 1, +i,
0)/\sqrt{2}$ in the Jackson convention, but become
$\epsilon_{K^*}^{\mu}(\pm 1)=(0, \pm 1, -i, 0)/\sqrt{2}$ in the
Jacob-Wick convention. Therefore in the NF, $\arg(\overline
A_{\parallel,\perp}/\overline A_0)$ equal to $\pi$ in the Jackson
convention, but are zero in the Jacob-Wick convention. Note that
in the two conventions, the longitudinal polarization vectors are
the same as $\epsilon_{\phi}^{\mu}(0)=(p_c, 0, 0, E_\phi)/m_\phi$
and $\epsilon_{K^*}^{\mu}(0)=(p_c, 0, 0, -E_{K^*})/m_{K^*}$.  Here
the amplitudes satisfy $\overline A =\overline A_0 +\overline
A_\parallel +\overline A_\perp, A = A_0 + A_\parallel - A_\perp$,
where the kinematic factors are not shown.}, they are given by
 \bea {\overline H_{00}} =
\frac{G_F}{\sqrt 2} (V_{tb} V_{ts}^*) a^0_{SM} (i f_\phi m_\phi)
(m_B + m_{K^*}) \left[a A_1(m_\phi^2) - b A_2(m_\phi^2)\right],
\nonumber \\
{\overline H_{\pm \pm}} =- \frac{G_F}{\sqrt 2} (V_{tb} V_{ts}^*)
a^\pm_{SM} (i f_\phi m_\phi) \left[(m_B + m_{K^*}) A_1(m_\phi^2)
\mp \frac{2 m_B p_c}{m_B + m_{K^*}} V(m_\phi^2) \right],
\label{eq:ampSM}
 \eea
with the constants  $a = (m_B^2 - m_\phi^2 - m_{K^*}^2)/(2 m_\phi
m_{K^*})$, $b = (2 m_B^2 p_c^2) /[m_\phi m_{K^*} (m_B +
m_{K^*})^2]$. Here $a^h_{SM} = a^h_3 + a^h_4 + a^h_5 -
\frac{1}{2}(a^h_7 + a^h_9 +a^h_{10})$. The superscript $h$ in
$a_i^h$'s denotes the polarization of $\phi$ and $\overline
K^{*0}$ mesons; $h=0$ is for the helicity~00 state and $h=\pm$ for
helicity~$\pm\pm$ states. Note that the weak phase effect is tiny
in $a^h_{SM}$ and is thus neglected in the study. Such helicity
dependent effective coefficients $a^h_{SM}$ do arise in the QCDF,
however in the naive factorization (NF), they turns out to be
same, i.e. $a^0_{SM} = a^+_{SM} = a^-_{SM} = a_{SM}$. In the NF,
one can rewrite the above amplitudes in the transversity basis as
 \bea \label{smamp}
{\overline A}_0^{SM} &=&\frac{G_F}{\sqrt 2} (V_{tb} V_{ts}^*)
a_{SM} ( i f_\phi m_\phi) (m_B + m_{K^*}) \left[ a A_1(m_\phi^2) -
b A_2(m_\phi^2)\right],
\nonumber \\
{\overline A}_\|^{SM} &=& -\frac{G_F}{\sqrt 2}(V_{tb} V_{ts}^*)
a_{SM} (i {\sqrt 2} f_\phi m_\phi) (m_B + m_{K^*}) A_1(m_\phi^2),
\nonumber \\
{\overline A}_\perp^{SM} &=&  -\frac{G_F}{\sqrt 2} (V_{tb}
V_{ts}^*) a_{SM} ({i \sqrt 2} f_\phi m_\phi) \frac{2 p_c m_B}{(m_B
+ m_{K^*})} V(m_\phi^2).
 \eea
 In the QCDF, $a_i^h$'s are
given by
 \bea \label{ai}
 a_1^h &=& c_1+
{c_2\over N_c}+{\alpha_s\over 4\pi}\,
 {C_F\over N_c}c_2\,(F^h+f_{II}^h),
\nonumber \\
 a_2^h &=& c_2+{c_1\over N_c}+{\alpha_s\over 4\pi}\,{C_F\over
N_c}c_1\,(F^h+f_{II}^h), \non \\
 a_3^h &=& c_3+{c_4\over N_c}+{\alpha_s\over
4\pi}\,{C_F\over N_c} c_4\,(F^h+f_{II}^h), \nonumber \\
 a_4^h &=& c_4+{c_3\over N_c}+{\alpha_s\over 4\pi}\,{C_F\over
N_c}\Bigg\{c_3\big[(F^h+f_{II}^h)+G^h(s_s)+G^h(s_b)\big]-c_1\left({\lambda_u\over
\lambda_t}G^h(s_u)+{\lambda_c\over\lambda_t}G^h(s_c)\right)
\nonumber
\\ && +(c_4+c_6) \sum_{i=u}^b \biggl(G^h(s_i)-{2\over3}\biggr)  +{3\over 2} (c_8+c_{10})\sum
_{i=u}^b e_i \biggl(G^h(s_i)-{2\over3}\biggr)\nonumber\\
&& +{3\over 2}c_9\big[e_q G^h(s_q)-{1\over
3}G^h(s_b)\big]+c_g G^h_g\Bigg\}, \non \\
 a_5^h &=& c_5+{c_6\over N_c}
 -{\alpha_s\over 4\pi}\,{C_F\over N_c} c_6(\tilde F^h + f_{II}^h+12),\non \\
 a_6^h &=& c_6+{c_5\over N_c}, \non \\
 a_7^h &=& c_7+{c_8\over N_c}-{\alpha_s\over 4\pi}\,{C_F\over N_c}
c_8(\tilde F^h+  f_{II}^h+12)-{\alpha\over 9\pi}\,N_c C^{h}_e,  \non \\
 a_8^h &=& c_8+{c_7\over N_c},  \non \\
 a_9^h &=& c_9+{c_{10}\over N_c}+{\alpha_s\over 4\pi}\,{C_F\over N_c}
 c_{10}\,(F^h+f_{II}^h)-{\alpha\over 9\pi}\,N_c C^{h}_e,  \non \\
 a_{10}^h &=& c_{10}+{c_9\over
N_c}+{\alpha_s\over 4\pi}\,{C_F\over N_c}
c_9\,(F^h+f_{II}^h)-{\alpha\over 9\pi}\,C^h_e, \eea
 where $C_F=(N_c^2-1)/(2N_c)$, $s_i=m_i^2/m_b^2$,
$\lambda_{q}= V_{qb}V^*_{qq'}$, and $q'=d,s$. Note that we have
given the expressions for $a_1^h$, $a_2^h$, which may be relevant
for charged $B$ decays, arising due to $O_1$ and $O_2$ in
Eq.~(\ref{eff}). There are QCD and electroweak penguin-type
diagrams induced by the 4-quark operators $O_i$ for
$i=1,3,4,6,8,9,10$. These corrections are described by the
penguin-loop function $G^h(s)$ given by
 \bea
 G^0(s) &=& {2\over 3}-{4\over 3}\ln{\mu\over m_b}+4\int^1_0
du\,\Phi^{\phi}_\|(u)\int^1_0 dx\,x(1-x)\ln[s-\bar u x(1-x)], \non \\
G^\pm(s) &=& {2\over 3}-{2\over 3}\ln{\mu\over m_b}\nonumber\\
&&+ 2\int^1_0 du\, \Bigg( g^{\phi(v)}_\perp(u) \pm {1\over 4}{d
g_\perp^{\phi(a)}(u) \over du }\Bigg) \int^1_0
dx\,x(1-x)\ln[s-\bar u x(1-x)]\,.
 \eea
 In Eq.~(\ref{ai}) we have also included the leading electroweak
penguin-type diagrams induced by the operators $O_1$ and $O_2$,
 \bea
  C_e^h &=& \left({\lambda_u\over \lambda_t}G^h(s_u)+{\lambda_c\over
\lambda_t}G^h(s_c)\right) \left(c_2+{c_1\over N_c}\right).
 \eea
 The dipole operator $O_{8g}$ will give a tree-level contribution
proportional to
\bea
&&  G^0_g = -2\int^1_0 du\,{\Phi^{\phi}_\|(u)\over 1-u}\,, \nonumber\\
&&  G^\pm_g =\int^1_0 {du\over \bar u}\,\Bigg[ \int_0^u\,
\Big(\Phi_\|^\phi(v)-g^{\phi(v)}_\perp(v)\Big)dv -\bar u
g_\perp^{\phi (v)}(u) \mp {{\bar u} \over 4}{d
g_\perp^{\phi(a)}(u) \over du} + {g_\perp^{\phi(a)}(u) \over
4}\Bigg]\,, \label{eq:cg}
 \eea
In Eq.~(\ref{ai}), the vertex correction is given by
 \bea
F^h=-12\ln{\mu\over m_b}-18+f_I^h, \label{F}
 \eea
where we have used the na\"{i}ve dimensional regularization (NDR)
scheme~\cite{Buras:1990fn},
\bea
 && \gamma_\mu \gamma_\nu \gamma_\lambda (1-\gamma_5) \otimes
 \gamma^\mu \gamma^\nu \gamma^\lambda (1-\gamma_5)
 =4(4-\varepsilon) \gamma_\mu (1-\gamma_5) \otimes \gamma^\mu (1-\gamma_5),\\
 && \gamma_\mu \gamma_\nu \gamma_\lambda (1-\gamma_5) \otimes
 \gamma^\lambda (1-\gamma_5) \gamma^\nu \gamma^\mu
 =4(1-2\varepsilon) \gamma_\mu (1-\gamma_5) \otimes \gamma^\mu (1-\gamma_5),\\
 && \gamma_\mu \gamma_\nu \gamma_\lambda (1-\gamma_5) \otimes
 \gamma^\mu \gamma^\nu \gamma^\lambda (1+\gamma_5)
 =4(1+\varepsilon) \gamma_\mu (1-\gamma_5) \otimes \gamma^\mu (1+\gamma_5),\\
 && \gamma_\mu \gamma_\nu \gamma_\lambda (1-\gamma_5) \otimes
 \gamma^\lambda (1+\gamma_5) \gamma^\nu \gamma^\mu
 =4(4-4\varepsilon) \gamma_\mu (1-\gamma_5) \otimes \gamma^\mu (1+\gamma_5),
 \eea
with $D=4-2\varepsilon$, and have adopted the $\overline {\rm MS}$ substraction.
An explicit calculation for $f_I^h$, arising from vertex
corrections, yields \bea
 f_I^0 &=& \int^1_0dx\,\Phi_\|^{\phi}(x)\left(3{1-2x\over 1-x}\ln
x-3i\pi\right), \non \\
f_I^\pm &=& \int^1_0dx\,\Bigg( g_\perp^{\phi (v)}(x)\pm {1\over 4}
\frac{d g_\perp^{\phi(a)}(x)}{dx} \Bigg) \left(3{1-2x\over 1-x}\ln
x-3i\pi\right). \label{fI} \eea
 The hard kernel $f_{II}^h$  for
hard spectator interactions, arising from the hard spectator
interactions with a hard gluon exchange between the emitted vector
meson and the spectator quark of the $B$ meson, have the
expressions:
 \bea f_{II}^0 &=& {4\pi^2\over
N_c}\,{2f_Bf_{K^*}m_{K^*}\over h_{0}}\int^1_0 d\rho\,
{\Phi^B_1(\rho)\over \drho}\int^1_0 d\eta
\,{\Phi^{K^*}_\|(\eta)\over \deta}\int^1_0 d\xi\,
{\Phi^{\phi}_\|(\xi)\over \xi}, \non \\
 f_{II}^\pm &=& - {4\pi^2\over N_c}\,{f_B f^{T}_{K^*}\over m_Bh_{\pm}}(1\mp
1)\int^1_0 d\rho\, {\Phi^B_1(\rho)\over \drho}\int^1_0 d\eta\,
 {\Phi^{K^*}_\perp(\eta)\over \deta^2}\int^1_0 d\xi\, \nonumber\\
 &&\ \ \ \times \Bigg[ 2 \Bigg( g_\perp^{\phi(v)}(\xi)
 - {1\over 4}\frac{d g_\perp^{\phi(a)}(\xi)}{d\xi}\Bigg)
  +
 \Bigg( \frac{1}{\xi} - \frac{1}{\bar \xi} \Bigg)
 \int_0^\xi dv (\Phi_\parallel^\phi(v) -g_\perp^{\phi (v)} (v))
 \Bigg] \non \\
  && +
{4\pi^2\over N_c}\,{2f_Bf_{K^*}m_{K^*}\over m_B^2h_{\pm}}\int^1_0
 d\rho\, {\Phi^B_1(\rho)\over\drho}\int^1_0
 d\eta\,
\Bigg(g^{K^*(v)}_\perp(\eta) \pm {1\over 4}
 {d g^{K^*(a)}_\perp(\eta)\over d \eta} \Bigg) \int^1_0 d\xi
\nonumber\\
&&\ \ \ \times \Bigg\{\frac{\deta+\bar \xi}{\deta^2 \bar\xi}
\Bigg(g_\perp^{\phi(v)}(\xi) \pm {1\over 4}{d
g_\perp^{\phi(a)}(\xi) \over d\xi} \Bigg) + \frac{1}{\deta^2 \xi}
 \int_0^\xi dv (\Phi_\parallel^\phi(v) -g_\perp^{\phi (v)} (v))\Bigg\},
 \label{fII2}
\eea
 where
 \bea \label{hel} h_{0} &=&
(m_B^2-m_{K^*}^2-m_{\phi}^2)(m_B+m_{K^*})A^{B{K^*}}_1(m_{\phi}^2)
-{4m_B^2p_c^2\over m_B+m_{K^*} }\,A^{B{K^*}}_2(m_{\phi}^2), \non \\
h_{\pm} &=& (m_B+m_{K^*})A^{B{K^*}}_1(m_{\phi}^2)\mp {2m_Bp_c\over
m_B+m_{K^*} }\,V^{B{K^*}}(m_{\phi}^2).
 \eea
Note that $\tilde F^h$  can be obtained from $F^h$ in
Eq.~(\ref{F}) with the replacement of $g_\perp^{\prime\, \phi (a)}
\to -g_\perp^{\prime\, \phi (a)}$. We will introduce a cutoff of
order $\Lambda_{\rm QCD}/m_b$ to regulate the infrared divergence
in $f_{II}$. Note also that we have corrected~\footnote{We
especially thank A. Kagan for pointing out that some terms in
$C_g^h$ may be missed in Ref.~\cite{CY}. We were therefore
motivated to recalculate the QCDF decay amplitudes.} the QCDF
results in Ref.~\cite{CY} which were done by Cheng and one of us
(K.C.Y.). The key point for the calculation is that one needs to
consider correctly the projection operator in the momentum space,
as discussed in Appendix~\ref{app:a}, which may explain the
difference with Ref.~\cite{Li:2003he}~\footnote{Our $G_g^+$ does
not agree with that obtained by Yang {\it et
al.}~\cite{Dariescu:2003tx}.}. In the calculation, we take the
asymptotic light-cone distribution amplitudes (LCDAs) for the
light vector mesons, and a Gaussian form for the $B$ meson wave
function.~\cite{CY}.
 We now make a SM estimate for various helicity amplitudes from a
power counting point of view. For $\overline B^{0} \to \phi
\overline K^{*0}$, the helicity amplitude ${\overline H_{00}}$
arising from the $(V-A)\otimes (V-A)$ operators is ${\cal O}(1)$,
since each of $\phi$ and $\overline K^{*0}$ mesons, with the quark
and antiquark being left and right-handed helicities,
respectively, requires no helicity flip. For ${\overline H_{--}}$,
the helicity flip for the $\bar s$ quark in the $\phi$ meson is
required, resulting in $m_\phi/m_b$ suppression for the amplitude.
Finally in ${\overline H_{++}}$ two helicity flips for $s$ quarks
are required, one in $\phi$ meson and the other in the $\overline
B^{0} \to {\overline K^{*0}}$ form factor transition, which cause
a suppression by $(m_\phi/m_b)(\bar\Lambda/m_b)$, where
$\bar\Lambda=m_B-m_b$. In a nutshell, the three helicity
amplitudes in the SM can be approximated as ${\overline H_{00}} :
{\overline H_{--}} : {\overline H_{++}} \sim {\cal O}(1): {\cal
O}(1/m_b) : {\cal O}(1/m_b^2) $. One should note that for the
CP-conjugated ${B^0} \rightarrow \phi K^{*0}$, the above result
is modified to be $H_{00}:H_{++}:H_{--} \sim {\cal O}(1):{\cal
O}(1/m_b): {\cal O} (1/m_b^2) $. The results extended to various
possible NP operators together with the SM operators will be shown
later in Table~\ref{tab:choose} and also illustrated in
Fig.~\ref{fig:tv}.

\subsection{New physics: hints from the BABAR and BELLE observations}

The large transverse $ \overline B \to  \phi \overline K^{*}$
fractions as have been observed by BELLE and
BABAR~\cite{Aubert:2003xc,Chen:2003jf,Aubert:2004xc,Abe:2004ku},
may hint a departure from the SM expectation for the longitudinal
one. Within the SM, the QCDF calculation~\cite{CY} yields
 \bea
  1 - R_0 = {\cal O}(1/m_b^2)
  = R_T \,,
  \eea
 where $R_0 =|\overline A_0|^2 /|\overline A_{tot}|^2$ and  $R_T = R_\| + R_\perp =
(|\overline A_\| |^2 + |\overline A_\perp |^2) /|\overline
A_{tot}|^2$. The observation of large $R_T$, as large as $50\%$,
may be possible to be accounted for in the SM, but here we are
considering the new physics alternatively~\footnote{In the PQCD
approach, annihilation contributions appear to be too small to
resolve the puzzle~\cite{Hnli}, but Li~\cite{Li:2004mp} has
recently argued that an decrease in one of the $B \to K^*$ form
factors could be helpful. Nevertheless, using the QCDF,
Kagan~\cite{Kagan} showed that the suppressed annihilations could
account for the observations with modest values for the BBNS
parameter $\rho_A$. See also the discussion after
Eq.~(\ref{eq:rheffai}).}. In the transversity basis the recent
experiments~\cite{Aubert:2004xc,Abe:2004ku} have shown that
 \bea |A_0|^2 (= |\overline H_{00}|^2) \approx
|\overline A_T|^2 (= |\overline A_{\|}|^2 + |\overline A_{\perp}|^2)\,,
 \eea
where $\overline A_{\|} = (\overline H_{++} + \overline
H_{--})/{\sqrt 2}$ and $\overline A_{\perp} = - (\overline H_{++}
- \overline H_{--})/ {\sqrt 2}$. One may need NP to explain such a
large $R_T$ ($\approx R_0$). A set of NP operators contributing to
the different helicity amplitudes like
 \bea
\overline H_{00}:\overline H_{--}:\overline H_{++} &\sim & {\cal
O}(1/m_b):{\cal O}(1):{\cal O}(1/m_b^2)\,, \label{eq:new1}
 \eea
or
 \bea
  {\overline H_{00}}:{\overline H_{--}}:{\overline H_{++}}
  &\sim & {\cal O}(1/m_b):{\cal O}(1/m_b^2):{\cal
O}(1)  \label{eq:new2}\,,
 \eea
 could resolve such polarization anomaly.
 Note  that $\overline H_{--}$ (in the former case)
and  $\overline H_{++}$ (in the latter case) are of ${\cal O}(1)$,
while $\overline H_{00}$ is always of ${\cal O}(1/m_b)$,
in contrast to the SM expectation. The detailed reason will be seen below.

\subsubsection{New physics operators}

\begin{figure}[tbh]
\vspace{0cm}
  \centerline{\epsfig{figure=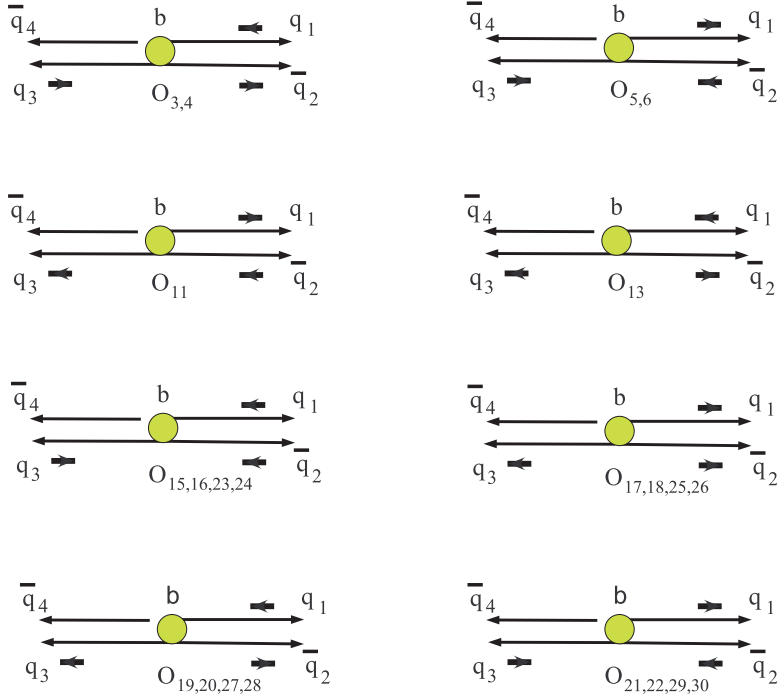,width=11cm}}
\vspace{0cm} \caption{\small\label{fig:tv} The main helicity
directions of quarks and antiquarks arising from various
four-quark operators during the $\overline B$ decay, where the
solid circle denotes the $b$ quark, $q_1,q_3$ are the $s$ quarks,
and $\bar q_2$ is the $\bar s$ quark. ($q_1, \bar q_2$) and ($q_3,
\bar q_4$) form $\phi$ and $\overline K^*$, respectively. $\bar
q_4$ is the spectator light quark which has no preferable
direction. The short arrows denote the helicities of quarks and
antiquarks. See the text for the detailed discussions.}
\end{figure}

Now Eqs.~(\ref{eq:new1}) and (\ref{eq:new2}) can serve as
guideline  in selecting NP operators. To begin with, consider the
following effective Hamiltonian ${\cal H}^{\rm NP}$:
 \bea
{\cal H}^{\rm NP}=\frac{G_F}{\sqrt{2}} \sum_{i=11}^{30}
\left[c_i(\mu) O_i (\mu)\right] + H.c.\,,
 \eea
which may be generated from some NP sources and contains the
following general NP four-operators:
 \begin{itemize}
 \item four-quark operators with vector and
axial-vector structures:
 \bea
 O_{11} &=& {\overline{s}} \gam^\m (1 + \gamma^5) b
 ~{\overline {s}}\gamma_\m(1 + \gam^5) s \,,
 \ \ \ \
O_{12} =  {\overline{s}_{\al}} \gam^\m(1 + \gamma^5) b_{\bt}
  ~{\overline {s}}_\bt  \gam_\m (1 + \gamma^5) s_\al \,,
\non \\
O_{13} &=&  {\overline{s}} \gam^\m(1 + \gamma^5) b
  ~{\overline {s}} \gam_\m(1 - \gamma^5) s \,,
 \ \ \ \
 O_{14} =  {\overline{s}_{\al}} \gam^\m(1 + \gamma^5) b_{\bt}
  ~{\overline {s}}_\bt  \gam_\m(1 - \gamma^5) s_\al \,,\nonumber\\
   \label{eq:vectorop}
 \eea
\item four-quark operators with scalar and pseudo-scalar
structures:
 \bea
O_{15} &=& {\overline{s}}(1 + \gamma^5) b ~{\overline {s}}(1 +
\gamma^5) s \,,
  \ \ \ \
O_{16} =  {\overline{s}_{\al}} (1 + \gamma^5)
b_{\bt}
  ~{\overline {s}}_\bt  (1 + \gamma^5) s_\al \,,
\non \\
O_{17} &=&  {\overline{s}} (1 - \gamma^5) b
  ~{\overline {s}} (1 - \gamma^5) s \,,
 \ \ \ \
O_{18} =  {\overline{s}_{\al}} (1 - \gamma^5) b_{\bt}
 s ~{\overline {s}}_\bt  (1 - \gamma^5) s_\al \,,
\non \\
O_{19} &=&  {\overline{s}}(1 + \gamma^5) b
  ~{\overline {s}}(1 - \gamma^5) s \,,
 \ \ \ \
 O_{20} =  {\overline{s}_{\al}} (1 + \gamma^5) b_{\bt}
  ~{\overline {s}}_\bt  (1 - \gamma^5) s_\al \,,
 \non \\
O_{21} &=&  {\overline{s}} (1 - \gamma^5) b
  ~{\overline {s}} (1 + \gamma^5) s \,,
 \ \ \ \
O_{22} = {\overline{s}_{\al}} (1 - \gamma^5) b_{\bt}
 ~{\overline {s}}_\bt  (1 + \gamma^5) s_\al \,,
 \label{eq:scalarop}
 \eea
\item four-quark operators with  tensor and axial-tensor
structures:
 \bea O_{23} &=&  {\overline{s}}\sigma^{\mu\nu} (1 + \gamma^5) b
  ~{\overline {s}}\sigma_{\mu\nu} (1 + \gamma^5) s \,,
 \ \ \ \
O_{24} =  {\overline{s}_{\al}}\sigma^{\mu\nu} (1 + \gamma^5)
b_{\bt}
  ~{\overline {s}}_\bt \sigma_{\mu\nu} (1 + \gamma^5) s_\al \,,
\non \\
O_{25} &=&  {\overline{s}}\sigma^{\mu\nu} (1 - \gamma^5) b
  ~{\overline {s}}\sigma_{\mu\nu} (1 - \gamma^5) s \,,
 \ \ \ \
O_{26} = {\overline{s}_{\al}}\sigma^{\mu\nu} (1 - \gamma^5)
b_{\bt}
  ~{\overline {s}}_\bt \sigma_{\mu\nu} (1 - \gamma^5) s_\al \,,
\non \\
 O_{27} &=&  {\overline{s}}\sigma^{\mu\nu} (1 + \gamma^5) b
  ~{\overline {s}}\sigma_{\mu\nu} (1 - \gamma^5) s \,,
 \ \ \ \
O_{28} =  {\overline{s}_{\al}}\sigma^{\mu\nu} (1 + \gamma^5)
b_{\bt}
  ~{\overline {s}}_\bt \sigma_{\mu\nu} (1 - \gamma^5) s_\al \,,
\non \\
O_{29} &=&  {\overline{s}}\sigma^{\mu\nu} (1 - \gamma^5) b
  ~{\overline {s}}\sigma_{\mu\nu} (1 + \gamma^5) s \,,
 \ \ \ \
O_{30} = {\overline{s}_{\al}}\sigma^{\mu\nu} (1 - \gamma^5)
b_{\bt}
  ~{\overline {s}}_\bt \sigma_{\mu\nu} (1 + \gamma^5) s_\al
  \,.\nonumber\\
   \label{eq:tensorop}
\eea
 \end{itemize}
 Here $c_{i}$ with $i=11,\dots,30$ are the Wilson
coefficients of the corresponding NP operators and $\mu$ the
renormalization scale, chosen to be $m_b$ here. Now we give an
estimation of several types of NP operators, contributing to
various ${\overline B^0} \to \phi \overline K^*$ helicity amplitudes. In
Fig.~\ref{fig:tv}, we draw the diagrams in the $\overline B$ rest
frame, where $q_1,q_3$ are the $s$ quarks, and $\bar q_2$ is the
$\bar s$ quark. ($q_1, \bar q_2$) and ($q_3, \bar q_4$) form
$\phi$ and $\overline K^*$, respectively. $q_4$ is the spectator light quark
which has not any preferable direction. $q_1, \bar q_2, q_3$ are
originated from the following operators: $\bar q_3 \Gamma_1 b \
\bar q_1 \Gamma_2 q_2$ for $O_{3-6}, O_{11-14}, O_{23-30}$ or
$\bar q_1 \Gamma_1 b \ \bar q_3 \Gamma_2 q_2$ for $O_{15-22}$. If
the helicity for $q_1$ or $\bar q_2$ is flipped, then the
amplitude is suppressed by a factor of $m_\phi /m_b$. On the other
hand, if the helicity of $q_3$ is further flipped, the amplitude
will be suppressed by $(m_\phi/m_b)(\bar\Lambda/m_b)$, with
$\bar\Lambda=m_B-m_b$. The results are summarized in
Table~\ref{tab:choose}.

\begin{table*}[ht]
\caption{Possible NP operators and their candidacy in satisfying
the anomaly resolution criteria. We have adopted the convention
$\Gamma_1 \otimes \Gamma_2 \equiv {\overline{s}} \Gamma_1 b \
{\overline {s}} \Gamma_2 s$.\label{tab:choose}}
\begin{ruledtabular}
\begin{tabular}{lccccr}
 Model & Operators & $\overline H_{00}$ & $\overline H_{--}$ & $\overline H_{++}$ & Choice\\
\hline
 SM & $ \gamma^\mu (1 - \gamma_5)\otimes \gamma_\mu (1 \mp \gamma_5 )$ &
${\cal O}(1)$ & ${\cal O}(1/m_b)$ & ${\cal O}(1/m_b^2)$ & \\
\hline
NP & $ \gamma^\mu (1 + \gamma_5)\otimes \gamma_\mu
(1 + \gamma_5 )$ & ${\cal O}(1)$ & ${\cal O}(1/m_b^2)$ & ${\cal O}(1/m_b)$ & N \\
\hline NP & $ \gamma^\mu (1 + \gamma_5)\otimes \gamma_\mu
(1 - \gamma_5 )$ & ${\cal O}(1)$ & ${\cal O}(1/m_b^2)$ & ${\cal O}(1/m_b)$ & N \\
\hline NP & $ (1 + \gamma_5)\otimes (1 + \gamma_5 )$ &
${\cal O}(1/m_b)$ & ${\cal O}(1)$ & ${\cal O}(1/m_b^2)$ & Y \\
\hline NP & $ (1 - \gamma_5)\otimes (1 - \gamma_5 )$ &
${\cal O}(1/m_b)$ & ${\cal O}(1/m_b^2)$ & ${\cal O}(1)$  & Y \\
\hline NP & $ (1 + \gamma_5)\otimes (1 - \gamma_5 )$ &
${\cal O}(1)$ & ${\cal O}(1/m_b^2)$ & ${\cal O}(1/m_b)$ & N \\
\hline
NP & $ (1 - \gamma_5)\otimes
(1 + \gamma_5 )$ &
${\cal O}(1)$ & ${\cal O}(1/m_b)$ & ${\cal O}(1/m_b^2)$ & N \\
\hline
NP & $\sigma^{\mu\nu} (1 + \gamma_5)\otimes \sigma_{\mu
\nu} (1 + \gamma_5 )$ &
${\cal O}(1/m_b)$  & ${\cal O}(1)$ & ${\cal O}(1/m_b^2)$ & Y \\
\hline
NP & $\sigma^{\mu\nu} (1 - \gamma_5)\otimes \sigma_{\mu
\nu} (1 - \gamma_5 )$ &
${\cal O}(1/m_b)$ & ${\cal O}(1/m_b^2)$ & ${\cal O}(1)$ & Y \\
\hline
 NP & $\sigma^{\mu\nu} (1 + \gamma_5)\otimes \sigma_{\mu
\nu} (1 - \gamma_5 )$ &
${\cal O}(1)$ & ${\cal O}(1/m_b^2)$ & ${\cal O}(1/m_b)$ & N \\
\hline
 NP & $\sigma^{\mu\nu} (1 - \gamma_5)\otimes \sigma_{\mu \nu} (1 + \gamma_5 )$
 & ${\cal O}(1)$ & ${\cal O}(1/m_b)$ & ${\cal O}(1/m_b^2)$ & N
\end{tabular}
\end{ruledtabular}
\end{table*}

From the Table~\ref{tab:choose}, we see that both (pseudo-)scalar
operators $O_{15-18}$ and (axial-)tensor operators $O_{23-26}$
satisfy the anomaly resolution criteria as given by
Eqs.~(\ref{eq:new1}) and (\ref{eq:new2}), while the rest are not.
However, through the Fierz transformation, it can be shown that
 $O_{15,16}$ and $O_{17,18}$  operators can be expressed
as a linear combination of $O_{23,24}$ and $O_{25,26}$ operators, respectively,
i.e.,
 \bea \label{eq:fiez}
 O_{15} &=& \frac{1}{12}O_{23}- \frac{1}{6} O_{24} ,\nonumber
  \\
 O_{16} &=& \frac{1}{12}O_{24}- \frac{1}{6} O_{23} ,\nonumber
 \\
  O_{17} &=& \frac{1}{12}O_{25}- \frac{1}{6} O_{26} ,\nonumber
  \\
 O_{18} &=& \frac{1}{12}O_{26}- \frac{1}{6} O_{25} .
 \eea
  Before we continue the
study, five remarks are in order. (i) The $\bar s \sigma^{\mu\nu}
(1+\gamma_5) b \ \bar s\sigma_{\mu\nu} (1+\gamma_5) s$ operator,
which could maintain $|\overline A_{\perp}|^2\approx |\overline A_{\|}|^2$,
was first
mentioned by Kagan~\cite{Kagan}~\footnote{However, the
contributions arising from the $\bar s \sigma^{\mu\nu}
(1+\gamma_5) b \ \bar s\sigma_{\mu\nu} (1-\gamma_5) s$ operator to
different polarization amplitudes should be $\overline
H_{00}:\overline H_{--}:\overline H_{++} \sim {\cal O}(1):{\cal
O}(1/m_b^2) :{\cal O}(1/m_b)$, not as mentioned in~\cite{Kagan}.}.
(ii) We do not consider NP of left-handed currents, $\bar s
\gamma_\mu(1-\gamma_5) b\ \bar s \gamma^\mu(1\mp\gamma_5) s$,
which give corrections to SM Wilson coefficients, $c_{1-10}$,
since they have no help for understanding large polarized
amplitudes and are strongly constrained by other $B\to PP, VP$
observations~\cite{hfag}. (iii) $O_{11-14}$ are the so-called
right-handed currents, emphasized recently by Kagan~\cite{Kagan}.
These operators give corrections to amplitudes as
 \bea
 \overline A_{0,\|} &\propto& (-V_{tb} V_{ts}^*) a_{SM} - (a_{11}+a_{12}+a_{13}), \nonumber\\
 \overline A_{\perp} &\propto& (-V_{tb} V_{ts}^*) a_{SM} + (a_{11}+a_{12}+a_{13}),
 \eea
where
 \bea\label{eq:rheffai}
 a_{11}=
 c_{11}+\frac{ c_{12}}{N_c}+{\rm nonfact.},\
 a_{12}=c_{12}+\frac{ c_{11}}{N_c}+{\rm nonfact.}, \
 a_{13}=  c_{13}+\frac{c_{14}}{N_c}+{\rm nonfact.},
\eea
 with ``nonfact."$\equiv$ nonfactorizable corrections.
  Note that $a_{11,12,13}$ enter the $\overline A_{0,\|}$
amplitudes with a ``minus" sign due to the relative sign changed
for $A_1, A_2$ form factors as compared to the SM amplitudes in
Eq.~(\ref{eq:ampSM}). If the right-handed currents contribute
constructively to $\overline A_\perp$ but destructively to
$\overline A_{0,\|}$, then one may have larger $|\overline A_\perp
/ \overline A_0 |^2$ to account for the data. According to the SM
result $|\overline A_\perp / \overline A_0 |^2\simeq 0.02$ in
Eq.~(\ref{eqn:smampnu1}), we need to have
$|a_{11}+a_{12}+a_{13}|/|-V_{tb} V_{ts}^* a_{SM}|\sim 1.5$ such
that $|\overline A_\perp / \overline A_0 |^2\simeq 0.25$. However
the resulting $|\overline A_{\|}|^2 \ll |\overline A_\perp|^2$
will be in contrast to the recent
observations~\cite{Aubert:2004xc,Abe:2004ku}. (iv) Since, in large
$m_b$ limit, the strong interaction conserves the helicity of a
produced light quark pair, helicity conservation requires that the
outgoing $s$ and $\bar s$ arising from $s-\bar s-n\ gluons$ vertex
have {\it opposite helicities}. The contribution of the
chromomagnetic dipole operator to the transversely polarized
amplitudes should be suppressed as $\overline H_{00}:\overline
H_{--}: \overline H_{++} \sim {\cal O}(1):{\cal O}(1/m_b): {\cal
O} (1/m_b^2)$; otherwise the results will violate the angular
momentum conservation. Actually if only considering
the two parton scenario for the meson, the contributions of
the chromomagnetic dipole operator to the transversely polarized amplitudes
equal to zero~\cite{Kagan} (see Eq.~(\ref{eq:cg}) and Appendix~\ref{app:a}
for the detailed discussions). Similarly, the $s$ and $\bar s$ quark pairs
generated from $c, \bar c$ annihilation in the charming penguin
always have {\it opposite helicities} due also to the helicity
conservation. Hence, that contributions to the transversely
polarized amplitudes are relative suppressed, in contrast with the
results in Refs.~\cite{Colangelo:2004rd,Cheng:2004ru}. (v) With
the same reason as the above discussion, in the SM, the
transversely polarized amplitudes originating from annihilations
are subjected to helicity suppression. A suggestion pointed out by
Kagan~\cite{Kagan} for the polarization anomaly is the
annihilation via the $(S-P)\otimes (S+P)$ operator, which
contributes to helicity amplitudes as ${\overline H}_{00},
{\overline H}_{--} \sim {\cal O}[(1/m_b^2) \ln^2
(m_b/\Lambda_{h})]$, ${\overline H}_{++} \sim {\cal O}[(1/m_b^4)
\ln^2 (m_b/\Lambda_{h})]$. However this contribution to $\overline
H_{--}$ is already of order $1/{m_b}^2$ although it is logarithmic
divergent.

 We now calculate the decay amplitudes for ${\overline
B^0} \rightarrow \phi {\overline K^{*0}}$ due to $O_{15-18}$ and
$O_{23-26}$ operators in Eqs.~(\ref{eq:scalarop}) and
(\ref{eq:tensorop}). The amplitudes for $B^0 \to \phi K^{*0}$ can
be obtained by $CP$-transformation. The
 matrix elements for (axial-)tensor operators $O_{23,25}$ can be recast into
\begin{eqnarray}
&& \langle \phi(q,\epsilon_1),\overline K^*(p',\epsilon_2)|
 \bar s \sigma^{\mu\nu}  (1\pm \gamma_5) s\
 \bar s \sigma_{\mu\nu} (1\pm \gamma_5) b | \overline B(p) \rangle
  = \left(1+\frac{1}{2N_c}\right) \nonumber\\
 && \ \  \times f_\phi^{T}\Bigg( 8
\epsilon_{\mu\nu\rho\sigma} \epsilon_1^{*\mu}\epsilon_2^{*\nu}
p^\rho p^{\prime\sigma} \, T_1(q^2)
   \mp 4i T_2(q^2) \left[ (\epsilon_1^*\cdot\epsilon_2^*)
  (m_B^2-m_{K^*}^2) - 2(\epsilon_1^*\cdot p) \,(\epsilon_2^* \cdot
  p)\right]
\nonumber\\
& &{ }\ \ \ \
 \pm 8 i T_3(q^2) (\epsilon_1^*\cdot p)
 (\epsilon_2^*\cdot p) \frac{m_\phi^2}{m_B^2-m_{K^*}^2} \Bigg)
\,, \label{eq:T}
\end{eqnarray}
under factorization, where the tensor decay constant $f_\phi^{T}$
is defined by~\cite{alisafir,Ali:2004hn,Ball:1998kk}
 \bea
  \langle\phi(q,\eps_1)|{\overline{s} \sigma^{\mu\nu} s}|0 \rangle
  = -i f_\phi^{T} (\eps_1^{\mu *} q^\nu - \eps_1^{\nu *} q^\mu ),
 \eea
 and
 \bea
 && \langle \overline K^*(p',\epsilon_2) | \bar s
\sigma_{\mu\nu} q^\nu (1+\gamma_5) b | \overline B(p)\rangle \nonumber\\
 &&= i\epsilon_{\mu\nu\rho\sigma} \epsilon^{*\nu} p^\rho
p^{\prime\sigma} \, 2 T_1(q^2)
 + T_2(q^2) \left\{ \epsilon^*_{2,\mu}
  (m_B^2-m_{K^*}^2) - (\epsilon^*\cdot p) \,(p+p')_\mu \right\}\nonumber\\
& & {} + T_3(q^2) (\epsilon_2^* \cdot p_B) \left\{ q_\mu -
\frac{q^2}{m_B^2-m_{K^*}^2}\, (p+p')_\mu \right\},
 \eea
 with
\begin{equation}
 T_1(0)  =  T_2(0). \label{eq:T1T2}
\end{equation}
The helicity amplitudes for the ${\overline B^0}$ decay due to the
NP operators are (in units of $G_F/\sqrt{2}$) given by
 \bea {\overline H}_{00}^{NP} &=& - 4 i f_\phi^{T} m_B^2 \left(
{\tilde a}_{23} - {\tilde a}_{25}\right)\left[h_2 T_2 (m_\phi^2) -
h_3 T_3(m_\phi^2)\right] \,,
\nonumber \\
{\overline H}_{\pm \pm}^{NP} &=& - 4 i f_\phi^{T} m_B^2 \left\{
{\tilde a}_{23} \left[ \pm f_1 T_1(m_\phi^2) - f_2 T_2(m_\phi^2)
\right] + {\tilde a}_{25} \left[ \pm f_1 T_1(m_\phi^2) + f_2
T_2(m_\phi^2) \right] \right\} \,,\nonumber\\
 \eea
and in the transversity basis, the amplitudes becomes (in units of
$G_F/\sqrt{2}$)
 \bea \label{npamp}
 {\overline A_{0}^{NP}} &=& -4 i f_\phi^{T}
m_B^2 \left[ {\tilde a}_{23} - {\tilde a}_{25}\right] \left[h_2
T_2(m_\phi^2) -  h_3 T_3(m_\phi^2)\right] ,
\non \\
{\overline A_{\|}^{NP}} &=&  4 i \sqrt{2} f_\phi^{T} m_B^2
({\tilde a}_{23} - {\tilde a}_{25})  f_2 T_2(m_\phi^2) ,
\non \\
{\overline A_{\perp}^{NP}} &=& 4 i \sqrt{2} f_\phi^{T} m_B^2
({\tilde a}_{23} + {\tilde a}_{25}) f_1 T_1 (m_\phi^2) ,
 \eea
where
 \bea \label{const}
 f_1 &=&
\frac{2p_c}{m_B}, ~ f_2 = \frac{m_B^2 - m_{K^*}^2}{m_B^2},~ ,
\non \\
h_2 &=& \frac{1}{2 m_{K^*} m_\phi}
\left(\frac{(m_B^2-m_\phi^2-m_{K^*}^2)(m_B^2-m_{K^*}^2)} {m_B^2} -
4p_c^2 \right),
\nonumber \\
h_3 &=& \frac{1}{2 m_{K^*} m_\phi} \left(\frac{4p_c^2 m_\phi^2}{
m_B^2 - m_{K^*}^2} \right),
 \eea
and
 \bea
 {\tilde a}_{23} &=&
 \left(1+ \frac{1}{2N_c} \right)
 \left( c_{23}+ \frac{1}{12} c_{15} - \frac{1}{6}
 c_{16} \right)
 + \left( \frac{1}{N_c}+\frac{1}{2} \right)\left( c_{24} + \frac{1}{12} c_{16} - \frac{1}{6}
 c_{15} \right) +{\rm nonfact.}, \nonumber\\
 {\tilde a}_{25} &=&  \left(1+ \frac{1}{2N_c} \right)\left( c_{25}+ \frac{1}{12} c_{17} - \frac{1}{6}
 c_{18} \right)
 +  \left( \frac{1}{N_c}+\frac{1}{2} \right)\ \left( c_{26} + \frac{1}{12} c_{18} - \frac{1}{6}
 c_{17} \right)+{\rm nonfact.} , \nonumber\\
  \eea
are NP effective coefficients defined by ${\tilde a}_{23} =
\left|{\tilde a}_{23}\right| e^{i\delta_{23}} e^{i\phi_{23}},
{\tilde a}_{25} = \left|{\tilde a}_{25}\right| e^{i\delta_{25}}
e^{i\phi_{25}}$ with $\phi_{23}, \phi_{25}$ being the
corresponding NP weak phases, while $\delta_{23}, \delta_{25}$ the
strong phases. Note that here we do not distinguish effective
coefficients for different helicity amplitudes since those
differences are relatively tiny compared with the hierarchy
results in Eqs.~(\ref{eq:new1}) and (\ref{eq:new2}). A further
model calculation for ${\tilde a}_{23}, {\tilde a}_{25}$ will be
published elsewhere~\cite{dy}. Note that if applying the equation
of motion to 4-quark operators in deriving the matrix in
Eq.~(\ref{npamp}), we can obtain the following relations: $T_1
\simeq V m_B/(m_B+m_{K^*}), T_2 \simeq A_1 m_B/(m_B-m_{K^*}), T_3
\simeq A_2$, consistent with results by the light-cone sum rule
(LCSR) calculation~\cite{alisafir,Ali:2004hn,Ball:1998kk}.
The $B^0 \to \phi K^{*0}$
polarization amplitudes can be obtained from the results of the
${\overline B^0} \to \phi \overline K^{*0}$ decay by performing
the relevant changes under CP-transformation. The total SM and NP
contributions for the $B^0$ and ${\overline B^0}$ decays can be
written as
 \bea \label{ampBbar}
   {\overline A}({\overline {B^0}} \rightarrow \phi {\overline K^{*0}})
   &=&
{\overline A}({\overline {B^0}} \rightarrow \phi {\overline
K^{*0}})_{SM} + {\overline A}({\overline {B^0}} \rightarrow \phi
{\overline K^{*0}})_{NP}, \nonumber\\
 A(B^0 \rightarrow \phi K^{*0}) &=& A(B^0
\rightarrow \phi K^{*0})_{SM} + A(B^0 \rightarrow \phi
K^{*0})_{NP}.
 \eea
With these decay amplitudes in the transversity basis, we can
evaluate physical observables: $|A_{0}|^2, |A_{\|}|^2,
|A_{\perp}|^2$,  $|{\overline A}_{0}|^2, |{\overline A}_{\|}|^2,
|{\overline A}_{\perp}|^2$, $ \Lambda_{00}, \Lambda_{\|\|},
\Lambda_{\perp\perp}, \Lambda_{\perp 0},
\Lambda_{\perp\|},\Lambda_{\| 0}, \Sigma_{00}, \Sigma_{\|\|},
\Sigma_{\perp\perp},\Sigma_{\perp 0},\Sigma_{\perp\|},\Sigma_{\|
0}$ and the triple products $A_T^0, A_T^{\|}$, $\overline A_T^0,
\overline A_T^{\|}$~\cite{Datta:2003mj}. The observables
$\Lambda_{hh}$ and $\Sigma_{hh}$ are defined as
 \bea \label{eq:def1} \Lambda_{hh} &=& \frac{1}{2}\left( | A_h
|^2 + | {\overline A}_h |^2 \right),
\nonumber \\
\Sigma_{hh} &=& \frac{1}{2}\left(| A_h |^2 - | {\overline A}_h |^2
\right),
\nonumber \\
\Lambda_{\perp i} &=& - Im(A_{\perp} A_i^* - {\overline
A_\perp}{\overline A_i^*}),
\nonumber \\
\Lambda_{\| 0} &=&  Re(A_{\|} A_0^* + {\overline A_\|}{\overline
A_0^*}),
\nonumber \\
\Sigma_{\perp i} &=& - Im(A_{\perp} A_i^* + {\overline
A_\perp}{\overline A_i^*}),
\nonumber \\
\Sigma_{\| 0} &=&   Re(A_{\|} A_0^* - {\overline A_\|}{\overline
A_0^*}),
 \eea
with $h = 0, \|, \perp$ and $i = 0,\|$. Here we adopt the normalization
conditions $\sum_i |\overline A_i|^2=1$ and $\sum_i |A_i|^2=1$.
The two triple products $A_T^0$ and $A_T^\|$ are defined as
 \bea \label{eq:def2}
 A_T^0 = \frac{Im (A_\perp
A_0^*)}{\left|A_0\right|^2 + \left|A_\| \right|^2 + \left|A_\perp
\right|^2},
\nonumber \\
A_T^\| = \frac{Im (A_\perp A_\|^*)}{\left|A_0\right|^2 + \left|A_\|
\right|^2 + \left|A_\perp \right|^2}.
 \eea
In our numerical analysis, we will focus on the studies of these
quantities.
  The CP-conjugated ${\overline A_T^0}$ and ${\overline A_T^\|}$
can be obtained by replacing $A_0, A_\|, A_\perp$ by their
CP-tranformed forms ${\overline A_0}, {\overline A_\|}, {\overline
A_\perp}$. Observables like $\Sigma_{\lam\lam}, \Sigma_{\| 0},
\Lambda_{\perp i}$ (with $\lam = 0, \|, \perp$ and $i = 0, \|$)
are sensitive to the NP~\cite{Datta:2003mj}, which, in absence of
the NP, strictly equal to zero. The triple product $A_T^0$ or
$A_T^\|$ can exhibit the relative phase between $A_\perp$ and
$A_0$ or between $A_\perp$ and $A_\|$. The differences between
$A_T^i$ and their CP-conjugated parts, i.e. $\Lambda_{\perp i} =
{\overline A_T^i} - A_T^i$ (with $i = 0, \|$), are CP-violating
(and also T-violating following from the CPT invariance theorem)
quantities. Therefore, any non-zero prediction of
$\Sigma_{\lam\lam}, \Sigma_{\| 0}, \Lambda_{\perp i}$ resembles
the evidence of a new source of CP-violation. Moreover, since
CP-violated effects are expected to be negligible within the SM,
sizable $\L_{\perp 0}$ or $\L_{\perp \|}$ may also imply the
existence of the NP. We will look for these possibilities from a
detailed numerical study.

\section{Input parameters}\label{sec:section3}

The decay amplitudes depend on the effective coefficients $a_i$'s,
KM matrix elements, several form factors, decay constants.

\subsection{KM matrix elements}

We will adopt the Wolfenstein parametrization, with parameters $A,
\lam, \rho$ and $\eta$, of the KM matrix as below
\[ V_{KM} =
\left( \begin{array}{ccc}
V_{ud} & V_{us}  & V_{ub} \\
V_{cd} & V_{cs}  & V_{cb} \\
V_{td} & V_{ts}  & V_{tb}\end{array} \right)
=\left(\begin{array}{ccc}
1 - \frac{1}{2}\lam^2 & \lam & A \lam^3 (\r - i \e) \\
-\lam & 1 - \frac{1}{2}\lam^2  & A \lam^2 \\
A \lam^3 (1 - \r- i \e) & - A \lam^2 & 1\end{array} \right).\]
 We employ $A$ and $\lam = \sin\theta_c$ at the values of $0.815$
and $0.2205$, respectively, in our analysis. The other parameters
are found to be $ {\overline \r} = \r (1 - \l^2/2) = 0.20 \pm
0.09$ and ${\overline \eta} = \eta (1 - \l^2/2) = 0.33 \pm 0.05$
\cite{PDG}.

\subsection{Effective coefficients $a_i$, form factors and decay constants}
The numerical values
for the effective coefficients $a_i^h$ with
$i=1,2,...,10$, which are obtained in the QCDF analysis~\cite{CY},
are cataloged in Table~\ref{tab:effectiveai}.  The effective
coefficients $a_i^0$ are the same both for $B$ and ${\overline
B}$, but not so for $a_i^{(-,+)}$. In the third and fifth columns
of Table~\ref{tab:effectiveai},  the $a_i$'s with the superscript
being bracketed are for the  $\overline B\to\phi \overline K^*$ process, otherwise for
the $B\to \phi K^*$ process.

\begin{table}[htbp]
\caption{Effective coefficients for $B\to \phi K^* ({\overline
B}\to \phi \overline K^*)$ obtained in the QCD factorization
analysis~\cite{CY}, where $a_{2,3}^{+,(-)}$, sensitive to the
nonfactorizable contribution $f_{II}^-$ which is opposite in sign to
$f_{II}^{0,+}$, are obviously different from $a_{2,3}^0,a_{2,3}^{-,(+)}$.
}\label{tab:effectiveai}
\begin{center}
\begin{tabular}{|c|c|c|c|c|c|c|}
\hline \hline $a^0_1$ & $1.0370+0.0135i$&$a^{+(-)}_1$
&$1.0900+0.0187i$ & $a^{-(+)}_1$&
$1.0180+0.0135i$ \\
\hline $a^0_2$ &$0.0764-0.0793i$ &$a^{+(-)}_2$ &$-0.2351-0.1096i$
& $a^{-(+)}_2$
&$0.1898-0.0793i$  \\
\hline
$a^0_3$ &$0.0055+0.0026i$ &$a^{+(-)}_3$ &$0.0156+0.0035i$ & $a^{-(+)}_3$&$0.0019+0.0026i$  \\
\hline
$a^0_4$ &$-0.0347-0.0068i$ &$a^{+(-)}_4$ &$-0.0366+0.001i$ & $a^{-(+)}_4$&$-0.0310-0.0036i$  \\
\hline
$a^0_5$ &$-0.0046-0.0030i$ &$a^{+(-)}_5$ &$0.0023-0.0019i$ & $a^{-(+)}_5$&$-0.0077-0.0030i$ \\
\hline $a^0_7$ &$-0.0001-0.0001i$ &$a^{+(-)}_7$ &$0.00001-0.0001i$
& $a^{-(+)}_7$&$0.00009
-0.00005i$ \\
\hline $a^0_9$ &$-0.0092-0.0003i$ &$a^{+(-)}_9$
&$-0.0096-0.0002i$ & $a^{-(+)}_9$&$-0.0088
-0.0002i$  \\
\hline $a^0_{10}$ &$-0.0004+0.0006i$ &$a^{+(-)}_{10}$
&$0.0023+0.0010i$ & $a^{-(+)}_{10}$&
$-0.0014+0.0007i$ \\
\hline
\end{tabular}
\end{center}
\end{table}
For the decay constants, we use~\cite{CY} $f_\phi = f_\phi^{T} =
237$~MeV, $f_{K^*}= f_{K^*}^T=160$~MeV, and $f_B=190$~MeV. For the
$B \to K^*$ transition form factors, we adopt the LCSR results
in~\cite{alisafir} with the parametrization
 \bea \label{eqn:fitform}
 F(q^2) = F (0)~ \exp\left[c_1 ({q^2/m_B^2}) + c_2
({q^2/m_B^2})^2\right]\,,
 \eea
which were rescaled to account for the $B\to K^* \gamma$ data. The
values of the relevant form factors and parameters are given in
Table~\ref{tab:table3}.
\begin{table}[htb]
\caption{\label{tab:table3} The values for the parametrization of
the $B \rightarrow K^*$ form factors in
Eq.~(\ref{eqn:fitform})~\cite{alisafir}. The renormalization scale
for $T_1, T_2, T_3$ is $\mu = m_b$.}
\begin{ruledtabular}
\begin{tabular}{cccccccc}
&$A_1$(0)&$A_2$(0)&$A_0$(0)&$V$(0)&$T_1$(0)&$T_2$(0)&$T_3$(0)\\
\hline
$F$(0)&0.294&0.246&0.412&0.399&0.334&0.334&0.234 \\
\hline
$c_1$&0.656&1.237&1.543&1.537&1.575&0.562&1.230 \\
\hline
$c_2$&0.456&0.822&0.954&1.123&1.140&0.481&1.089 \\
\end{tabular}
\end{ruledtabular}
\end{table}
The reason for choosing this set of form factors is because the
$T_1(0)$ value extracted from the $B\to K^* \gamma$ and $B \to X_s
\gamma$ data seems to prefer a smaller
one~\cite{Beneke:2001at,Bosch:2001gv,Ali:2004hn}.

\section{Numerical Analysis}\label{sec:section4}
We will estimate the NP parameters which may resolve the
polarization anomaly in $B({\overline B}) \rightarrow \phi
K^{*}({\overline K^*})$ decays~\cite{Aubert:2004xc,Abe:2004ku}. An
enhancement in transversely polarized amplitudes by $50 \%$ can
therefore take place in our NP scenario since the SM polarization
amplitudes, ${\overline A}_0^{SM}: {\overline A}_\|^{SM}
:{\overline A}_\perp^{SM} \sim {\cal O}(1): {\cal O}(1/m_b): {\cal
O}(1/m_b)$, are modified to be ${\overline A}_0^{NP} : {\overline
A}_\|^{NP} :{\overline A}_\perp^{NP} \sim {\cal O} (1/m_b): {\cal
O} (1): {\cal O}(1)$, as given in Eq.~(\ref{npamp}) which allows
us to find solutions in the NP parameter space ($\left| {\tilde
a}_{23}\right|, \delta_{23}, \phi_{23},\left| {\tilde
a}_{25}\right|, \delta_{25}, \phi_{25}$) for explaining the $B \to
\phi K^*$ polarization anomaly.

Choosing the normalization conditions $\sum_i |A_i|^2=\sum_i
|\overline A_i|^2= 1$, and setting $\arg(A_0)=\arg(\overline
A_0)=0$, one can measure the magnitudes and relative phases of the
six $A_{0,\|,\perp}, \overline A_{0,\|,\perp}$ polarization
amplitudes~\footnote{For simplicity, in the present study we have
chosen the convention $\arg(\overline A_0)=\arg(A_0)=0$, i.e. we
do not consider here the physics arising from the difference
between $\arg(\overline A_0)$ and $\arg(A_0)$.}, giving 8
measurements, and then extracts 12 observables in
Eq.~(\ref{eq:def1}) as well as the triple products in
Eq.~(\ref{eq:def2}). We take the average of the BABAR and BELLE
data in our $\chi^2$ analyses for estimating NP parameters and
consequently obtain the predictions for observables. For
simplicity, we neglect the correlations among the data. The
$\chi^2_i$ of any observable ${\cal O}_i$ with the measurement
${\cal O}_i(expt) \pm (1\sigma_i)_{expt}$ is defined as
 \bea
\label{eqn:chiind} \chi_i^2 = \left[\frac{{\mathcal O}_i (expt) -
{\mathcal O}_i(theory)} {(1\sigma_i)_{expt}}\right]^2. \eea For
$N$ different observables, the total $\chi^2$ equals to $\chi^2 =
\sum_{i = 1}^{N} \chi_i^2$. In the $\chi^2$ best fit analysis, we
consider the following $8$ observables:
  \bea \label{eqn:input}
  && |A_0|^2, |A_\perp|^2,  |\overline A_0|^2, |\overline
  A_\perp|^2, \arg(A_{\|}), \arg(A_{\perp}),
 \arg({\overline A}_{\|}), \arg({\overline A}_{\perp}),
  \eea
as our inputs. For the purpose of performing the numerical
analysis easily, we have converted the BABAR measurements into the
above quantities, as shown in Tables~\ref{tab:1st} and
\ref{tab:2nd}.

Since the interference terms in the angular distribution
analysis~\cite{Aubert:2004xc,Abe:2004ku} are limited to
$Re({\overline A}_\| {\overline A}_0^*)$, $Im({\overline A}_\perp
{\overline A}_0^*)$, and $Im({\overline A}_\perp {\overline
A}_\|^*)$, there exists a phase ambiguity:
 \bea \label{ambig}
\arg({\overline A}_\|) &\rightarrow& - \arg({\overline A}_\|),
\non \\
\arg({\overline A}_\perp) &\rightarrow&  \pm \pi - \arg({\overline
A}_\perp),
\non \\
\arg({\overline A}_\perp) - \arg({\overline A}_\|) &\rightarrow&
\pm \pi -(\arg({\overline A}_\perp) - \arg({\overline A}_\|)).
 \eea
Therefore, the world averages for $\arg({\overline A}_\|)$ and
$\arg({\overline A}_\perp)$, given in Tables~\ref{tab:1st} and
\ref{tab:2nd}, can be
 \bea
 \arg({\overline A}_\|) = -2.33 \pm 0.22, ~~
 \arg({\overline A}_\perp) = 0.59 \pm 0.24,
  \label{eqn:fold1}
 \eea
or, following from Eq.~(\ref{ambig}),
 \bea
 \arg({\overline A}_\|)   = 2.33 \pm 0.22, ~~
 \arg({\overline A}_\perp) = 2.55 \pm 0.24.
  \label{eqn:fold2}
 \eea
From Eq.~(\ref{eqn:fold1}), the phase difference for ${\overline
A}_\perp$ and ${\overline A}_\|$ reads
 \bea \arg({\overline
A}_\perp) - \arg({\overline A}_\|) \approx \pi, \label{eq:s1}
 \eea but, on the
other hand, from Eq.~(\ref{eqn:fold2}), becomes
 \bea
\arg({\overline A}_\perp) - \arg({\overline A}_\|) \approx 0. \label{eq:s2}
 \eea
The resultant implications in Eqs.~(\ref{eq:s1}) and (\ref{eq:s2})
are discussed below. Numerically, SM and NP amplitudes in the
transversity basis for ${\overline B^0} \rightarrow \phi
{\overline K^{*0}}$ are given by
 \bea \label{eqn:smampnu1} {\overline A}_0^{SM} &=& -0.00307 -
0.00074~i,
\non \\
{\overline A}_\|^{SM} &=& +0.00048 - 0.000064~i,
\non \\
{\overline A}_\perp^{SM} &=& +0.00040 - 0.000073 ~i,
 \eea  and
 \bea
\label{eqn:npampnu2} {\overline A}_0^{NP} &\simeq&
2.2\left(|{\tilde a}_{23}|e^{i(\delta_{23} + \phi_{23})} -
|{\tilde a}_{25}|e^{i(\delta_{25} + \phi_{25})}\right),
\non \\
{\overline A}_\|^{NP} &\simeq& -11.9\left(|{\tilde
a}_{23}|e^{i(\delta_{23} + \phi_{23})} - |{\tilde
a}_{25}|e^{i(\delta_{25} + \phi_{25})}\right),
\non \\
{\overline A}_\perp^{NP} &\simeq& -10.9\left(|{\tilde
a}_{23}|e^{i(\delta_{23} + \phi_{23})} + |{\tilde
a}_{25}|e^{i(\delta_{25} + \phi_{25})}\right), \eea respectively,
in units of $-i G_F/\sqrt{2}$. From Eq.~(\ref{eqn:smampnu1}), we
find that $|{\overline A}_\|^{SM}|/|{\overline A}_0^{SM}|$ and
$|{\overline A}_\perp^{SM}|/|{\overline A}_0^{SM}|$ are $0.17$ and
$0.14$, respectively. In other words, ${\overline
A}_{\|,\perp}^{SM}$ are ${\cal O}(1/m_b)$ suppressed, compared to
${\overline A}_0^{SM}$. On the other hand, the measurements for
$|\overline A_{\|} |^2 \approx |\overline A_\perp |^2 \approx
\frac{1}{2} |\overline A_0 |^2$, as cataloged in
Table~\ref{tab:1st}, mean that ${\overline A}_\|$ and ${\overline
A}_\perp$ are dominated by ${\overline A}_\|^{NP}$ and ${\overline
A}_\perp^{NP}$, respectively. We therefore find that the data for
the amplitude phases in Eq.~(\ref{eqn:fold1}) prefer the ${\tilde
a}_{25}$ terms in ${\overline A}_\|^{NP}$ and ${\overline
A}_\perp^{NP}$ given in Eq.~(\ref{eqn:npampnu2}), since there is a
phase difference of $\pi$ between two ${\tilde a}_{25}$ terms.
Consequently, from Table~\ref{tab:choose}, we get ${\overline
H}_{++} \gg {\overline H}_{--}$ if $O_{17,18},O_{25,26}$ NP
operators are dominant. On the other hand, for the data of the
amplitude phases in Eq.~(\ref{eqn:fold2}), we find that the
${\tilde a}_{23}$ terms in ${\overline A}_\|^{NP}$ and ${\overline
A}_\perp^{NP}$ in Eq.~(\ref{eqn:npampnu2}) are instead favored,
since they have the same sign. Accordingly, also from
Table~\ref{tab:choose}, as only $O_{15,16},O_{23,24}$ NP operators
are considered we obtain ${\overline H}_{--} \gg {\overline
H}_{++}$, which is consistent with the SM
expectation~\cite{Suzuki:2001za,Cheng:2001ez}.
Therefore because of the phase ambiguity,  the data prefer two
different types of NP scenarios: (i) the first scenario, where the
NP is characterized by $O_{17,18,25,26}$ operators, while the
operators $O_{15,16,23,24}$ are absent, (ii) the second scenario,
where the NP is dominated by $O_{15,16,23,24}$ operators, while
$O_{17,18,25,26}$ operators are absent.

\subsection {The first scenario with $O_{15,16,23, 24}$ absent}

In this scenario, the NP effects characterized by
$O_{17,18,25,26}$ operators are lumped into the single effective
coefficient ${\tilde a}_{25}=|\tilde a_{25}| e^{i\phi_{25}}
e^{i\delta_{25}}$, where $\phi_{25}$ and $\delta_{25}$ are the NP
weak and strong phases associated with it. Therefore, in our
$\chi^2$ analysis, we have three fitted parameters, $|{\tilde
a}_{25}|, \phi_{25}$, and $\delta_{25}$. The $\chi^2_{min}/d.o.f.$
for this scenario is $4.15/5$, where $d.o.f.\equiv$~degrees of
freedom in the fit. Our best fit results together with the data
are cataloged in Tables~\ref{tab:1st} and \ref{tab:observables}.
For illustration, we obtain theoretical errors by scanning the
$\chi^2\leq \chi^2_{\rm min.}+1$ parameter space. The BRs are only
sensitive to the form factors, while the rest results depend  very
weakly on the theoretical input parameters and the cutoff that
regulates the hard spectator effects in the SM calculation.  To
estimate the errors for BRs, arising from the input parameters, we
allow 10\% variation in form factors and decay constants which may
be underestimated, and the resulting errors are displayed in
Table~\ref{tab:observables}. The NP parameters are given by
 \bea \label{eqn:fit1}
  \left|{\tilde a}_{25}\right| = (2.10^{+0.19}_{-0.12})\times 10^{-4},
 \delta_{25}= 1.15\pm 0.09, \phi_{25} = -0.12\pm 0.09,
 \eea
with the phases in radians. Note that the non-small $\delta_{25}$
may imply that the strong phase due to annihilation mechanism in
the SM cannot be negligible.
{\squeezetable
\begin{table}[tb]
\caption{\label{tab:1st} Comparison between the first NP scenario
predictions and the average for BABAR and BELLE
data~\cite{Aubert:2004xc,Abe:2004ku} with the phases given in
Eq.~(\ref{eqn:fold1}). The $\chi_{min}^2/d.o.f.$ for $8$ inputs is
$4.15/5$.}
\begin{ruledtabular}
\begin{tabular}{l|lll|l}
NP parameters      & & & & NP results \\
\hline
$|{\tilde a}_{25}|$& & & & $(2.10^{+0.19}_{-0.12})\times 10^{-4}$ \\
$\delta_{25}$      & & & &  $1.15\pm 0.09$  \\
$\phi_{25}$        & & & & $ -0.12\pm 0.09$\\
\hline\hline
Observables & BABAR & BELLE & Average& NP results\\
\hline
  $\arg({\overline A}_\|)$ & $-2.61\pm 0.31$
 &$-2.05 \pm 0.31$ &$-2.33\pm 0.22 $ &$-2.60\pm 0.14$\\
  $\arg(A_\|)$ &$-2.07\pm 0.31$
 &$-2.29 \pm 0.37$ &$-2.16\pm 0.24$ &$-2.40\pm 0.14$\\
  $\arg({\overline A}_\perp)$ &$0.31\pm 0.36$
 &$0.81\pm 0.32$   &$0.59\pm 0.24$ & $0.87\pm 0.12$\\
  $\arg(A_\perp)$ &$1.03\pm 0.36$
 &$0.74\pm 0.33$   &$0.87\pm 0.24$ &$1.10\pm 0.13$\\
\hline
  $ |\overline A_0|^2, (|A_0|^2)$ &$0.49\pm0.07\ (0.55\pm0.08)$
 & $0.59\pm0.1\ (0.41\pm0.10)$    &$0.52\pm0.06~(0.50\pm0.06)$
 & $0.52\pm 0.04\ (0.53\pm 0.04)$\\
  $ |\overline A_{\|}|^2, (|A_{\|}|^2)$ & & &
 &$0.20\pm 0.02\ (0.22\pm 0.02)$ \\
  $ |\overline A_{\perp}|^2, (|A_{\perp}|^2)$
 &$0.20\pm0.07\ (0.24\pm 0.08)$   &$0.26\pm0.09\ (0.24\pm 0.10)$
 &$0.22\pm0.06~(0.24\pm0.07)$     &$0.28\pm0.03\ (0.25\pm 0.02)$ \\
\hline
 ${\overline A}_T^0, (A_T^0)$     & &$0.28 \pm 0.10~ (0.21\pm 0.09)$
 &                                &$0.19\pm 0.04~(0.24\pm 0.03)$\\
 ${\overline A}_T^\|,(A_T^\|)$    &
 &$0.06\pm 0.06~(0.04 \pm 0.08)$  & &$0.09\pm 0.01~(0.10\pm 0.00)$
\end{tabular}
\end{ruledtabular}
\end{table}
}
In Tables~\ref{tab:1st} and \ref{tab:observables}, we obtain
results in good agreement with the data. The BABAR and BELLE
\cite{Abe:2004ku,Aubert:2004xc} data show that \bea
\label{eqn:LDA1} \Lambda_{00} &\simeq& \Lambda_{\|\|} +
\Lambda_{\perp\perp},
\non \\
\Lambda_{\|\|} &\simeq& \Lambda_{\perp\perp}, \eea which can be
realized as follows. The transverse amplitudes are given by
 \bea \label{eqn:hpm1}
  {\overline A}_{\|} &=& {\overline A}_{\|}^{NP}+{\overline
  A}_{\|}^{SM},
 \non \\
 {\overline A}_{\perp} &=&  {\overline A}_{\perp}^{NP}+ {\overline
 A}_{\perp}^{SM},
 \eea
where ${\overline A}_{\|}^{NP}\approx -1.1{\overline
A}_{\perp}^{NP}$ in this scenario. In ${\overline A}_{\|}$ of
Eq.~(\ref{eqn:hpm1}), the interference of ${\overline
A}_{\|}^{NP}$ and ${\overline A}_{\|}^{SM}$ is destructive, while
in ${\overline A}_\perp$ the interference of ${\overline
A}_{\perp}^{NP}$ and ${\overline A}_{\perp}^{SM}$ becomes
constructive. We thus find $\Lambda_{\perp\perp} > \Lambda_{\|\|}$
and accordingly $\Lambda_{\|\|} \simeq 0.8 \Lambda_{\perp\perp}$.
Interestingly, because $\delta_{25}+\phi_{25}$ is much closer to
$0$ as compared to $\delta_{25}-\phi_{25}$, the above interference
effects thus result in $|\overline A_{\|}|^2/|\overline A_\perp|^2
< |A_{\|}|^2/|A_\perp|^2$. In other words, a larger $|\phi_{25}|$
yields larger magnitudes of $A_{CP}^\|, A_{CP}^\perp$. To get the
first relation of Eq.~(\ref{eqn:LDA1}), we first take the squares
of the ${\overline A}_{\|}$ and ${\overline A}_{\perp}$ of
Eq.~(\ref{eqn:hpm1}), and then add them up together with their
CP-conjugated parts. The interference terms are mutually cancelled
and one thus finds $\Lambda_{\|\|} + \Lambda_{\perp\perp} \sim
\Lambda_{00}$, due to $|{\overline A}_{\|,\perp}^{NP}|^2 \gg
|{\overline A}_{\|,\perp}^{SM}|^2$.

We obtain  $\Lambda_{\| 0}=-0.33\pm 0.04$ and $\Sigma_{\perp
0}=-0.44\pm 0.05$, as compared with the BELLE data: $\Lambda_{\|
0}= -0.39 \pm 0.14$ and $\Sigma_{\perp 0} = -0.49 \pm 0.14$.
Within the SM, $\Lambda_{\| 0}\simeq 0.30$ and $\Sigma_{\perp
0}\simeq 0.10$ are in contrast to the data.

The consistency between data and this NP scenario requires the
presence of a large strong phase $\delta_{25}$ and a (small) weak
phase $\phi_{25}$. Our numerical predictions for the rest NP
related observables are $\Lambda_{\perp\|},\Sigma_{00},
\Sigma_{\|\|}, \Sigma_{\perp\perp} \sim \pm (1-2)\%$, which are
marginal sensitive to $\phi_{25}$. Since our analysis yields
$\overline A_T^\|\simeq 0.09$ and $A_T^\|\simeq 0.10$, we
therefore obtain $\Sigma_{\perp \|}= - \overline A_T^\| -
A_T^\|\simeq -0.19$ and $\Lambda_{\perp \|}=\overline A_T^\|-
A_T^\|\simeq -0.01$. We observe that $\Lambda_{\perp 0,\perp \|},
\Sigma_{\perp\perp}, A_{CP}(\overline B\to \phi \overline K^*),
A_{CP}^{0,\|}$ have the same sign as $\phi_{25}$, whereas
$\Sigma_{00,\| \|, \| 0}, A_{CP}^{\perp}$ are of the opposite
sign. For a small $\phi_{25}=-0.12\ (=-7^\circ)$, we get NP
related quantities $\Lambda_{\perp 0}\approx -\Sigma_{\| 0}\approx
A_{CP}^\|\approx -0.05$ which may become visible in the future
once the experimental errors go down. It is interesting to note
that the existence of the non-zero NP weak phase $\phi_{25}$ may
be hinted by the BABAR measurements of $\arg(\overline A_\perp -
A_\perp)\neq 0$ and $\arg(\overline A_\| - A_\|)\neq 0$. If taking
alone the BABAR data as inputs, we obtain $\phi_{25} =
(16\pm7)^{\circ}$, which could cause sizable effects in
observations: $\Sigma_{\| 0} (=Re(A_{\|}A_0^*-\overline A_{\|}
\overline A_0^*))= 0.15\pm 0.07$, $A_T^0 (=0.26\pm0.03) \neq
\overline A_T^0 (=0.11\pm 0.06$), $\Sigma_{\perp 0} (=-\overline
A_T^0 - A_T^0) = -0.38\pm 0.08$, $\Lambda_{\perp 0} (=\overline
A_T^0- A_T^0)=-0.15\pm0.07$ and $A_{CP}^\| \simeq -2 A_{CP}^\perp
\simeq 3 A_{CP}^0 \simeq (-15\pm 6)\%$. As for the branching
ratio, we obtain BR($B^0 \to \phi K^{*0} )\simeq (1.33\pm 0.25)
\times 10^{-6} $, in good agreement with the world average
$(9.5\pm 0.9)\times 10^{-6}$~\cite{hfag}, while without NP
corrections the result becomes a much smaller value of $\sim
5.8\times 10^{-6}$.

\subsection {The second scenario with $O_{17,18,25,26}$ absent}
{\squeezetable
\begin{table}[tb]
\caption{\label{tab:2nd} Comparison between the second NP scenario
predictions and the average for BABAR and BELLE
data~\cite{Aubert:2004xc,Abe:2004ku} with the phases given in
Eq.~(\ref{eqn:fold2}). The $\chi_{min}^2/d.o.f.$ for $8$ inputs is
$0.56/5$.}
\begin{ruledtabular}\begin{tabular}{l|lll|l}
NP parameters      & & & & NP results \\
\hline
$-|{\tilde a}_{23}|$& & & & $-(1.70^{+0.11}_{-0.07})\times 10^{-4}$ \\
$\tilde\delta_{23}=\delta_{23}-\pi$      & & & &  $-0.78\pm 0.10$  \\
$\phi_{23}$        & & & & $ 0.14\pm 0.09$\\
\hline\hline
Observables & BABAR & BELLE & Average& NP results\\
\hline
  $\arg({\overline A}_\|)$&$2.61\pm 0.31$
 &$2.05 \pm 0.31$ &$2.33\pm 0.22 $ &$2.42\pm 0.17$\\
  $\arg(A_\|)$ &$2.07\pm 0.31$
 &$2.29 \pm 0.37$ &$2.16\pm 0.24$ &$2.21\pm 0.18$\\
  $\arg({\overline A}_\perp)$ &$2.83\pm 0.36$
 &$2.33\pm 0.32$   &$2.55\pm 0.24$ & $2.44\pm 0.17$\\
  $\arg(A_\perp)$ &$2.11\pm 0.36$
 &$2.40\pm 0.33$   &$2.27\pm 0.24$ &$2.24\pm 0.17$\\
\hline
  $ |\overline A_0|^2, (|A_0|^2)$ &$0.49\pm0.07\ (0.55\pm0.08)$
 & $0.59\pm0.1\ (0.41\pm0.10)$    &$0.52\pm0.06~(0.50\pm0.06)$
 & $0.51\pm 0.04\ (0.52\pm 0.04)$\\
  $ |\overline A_{\|}|^2, (|A_{\|}|^2)$ & & &
 &$0.26\pm 0.02\ (0.26\pm 0.02)$ \\
  $ |\overline A_{\perp}|^2, (|A_{\perp}|^2)$
 &$0.20\pm0.07\ (0.24\pm 0.08)$   &$0.26\pm0.09\ (0.24\pm 0.10)$
 &$0.22\pm0.06~(0.24\pm0.07)$     &$0.23\pm0.02\ (0.23\pm 0.02)$ \\
\hline
 ${\overline A}_T^0, (A_T^0)$     & &$0.28 \pm 0.10~ (0.21\pm 0.09)$
 &                                &$0.22\pm 0.04~(0.27\pm 0.04)$\\
 ${\overline A}_T^\|,(A_T^\|)$    &
 &$0.06\pm 0.06~(0.04 \pm 0.08)$  & &$0.01^{+0.00}_{-0.01}~(0.01^{+0.00}_{-0.01})$
\end{tabular}
\end{ruledtabular}
\end{table}
}

In the second scenario, the NP is characterized by
$O_{15,16,23,24}$ operators and the only relevant NP parameter is
${\tilde a}_{23}=|{\tilde a}_{23}|e^{i\phi_{23}} e^{i\delta_{23}}$
with $\phi_{23}$ and $\delta_{23}$ being the NP weak and strong
phases, respectively. Following the same way as in the first
scenario, we show the results in Tables~\ref{tab:2nd} and
\ref{tab:observables}, where $\chi^2_{min}/d.o.f.$ is $0.56/5$.
The NP parameters in this scenario are given by~\footnote{It may
be better to rewrite as ${\tilde a}_{23}=-|{\tilde
a}_{23}|e^{i\phi_{23}} e^{i\tilde\delta_{23}}$, where the
redefined strong phase is $\tilde\delta_{23}=\delta_{23}-\pi
=-0.78\pm 0.10 [=(-45\pm 6)^\circ]$. The reason is that it is hard
to have a large strong phase in the perturbation calculation.}
\bea \label{fit2}
 \left|{\tilde a}_{23}\right| = (1.70^{+0.11}_{-0.07})\times 10^{-4},
  \delta_{23}= 2.36\pm 0.10, \phi_{23} = 0.14\pm 0.09,
\eea with phases in radians. {\squeezetable
\begin{table}[tb]
\caption{\label{tab:observables} Comparison between the  NP
predictions and data~\cite{Aubert:2004xc,Abe:2004ku}. The NP
related observables are denoted by $``(*)"$. The second errors for
BRs come from the uncertainties of input parameters, and the first
ones are obtained with the constraint $\chi^2\leq \chi^2_{\rm
min.}+1$. The world average for total BR is $(9.5\pm 0.9)\times
10^{-6}$~\cite{hfag}.}
\begin{ruledtabular}
\begin{tabular}{l|ll|ll}
 &BABAR & BELLE & The 1st scenario & The 2nd scenario \\
 \hline\hline
 $\Lambda_{00}$                   & &$\ \ 0.50 \pm 0.07$
 & $\ \ 0.53\pm 0.04$ & $\ \ 0.51\pm 0.04$ \\
 $\Lambda_{\|\|}$                 & &$\ \ 0.25 \pm 0.07$
 & $\ \ 0.21\pm 0.02$ & $\ \ 0.26\pm 0.02$\\
 $\Lambda_{\perp\perp}$           & &$\ \ 0.25 \pm 0.07$
 & $\ \ 0.26\pm 0.02$ & $\ \ 0.23\pm 0.02$\\
 $\Lambda_{\| 0}$                 & &$-0.39 \pm 0.14$
 & $-0.33\pm 0.04$& $-0.49 \pm 0.07$\\
 $\Sigma_{\perp 0}$               & &$-0.49 \pm 0.14$
 & $-0.44\pm 0.05$&$-0.49 \pm 0.07$\\
 $\Sigma_{\perp \|}$              & &$-0.09 \pm 0.10$
 & $-0.19\pm 0.01$&$-0.01\pm 0.00$\\
 $\Lambda_{\perp 0}$(*)           &$-0.22\pm 0.10$& $ 0.07 \pm 0.12 $
                            & $-0.05\pm 0.04$ &$-0.05\pm 0.06$\\
 $\Lambda_{\perp \|}$(*)          &$\ \ 0.04\pm 0.08$&$0.02 \pm 0.10 $
                            &$-0.01^{+0.00}_{-0.01}$ &$-0.00^{+0.00}_{-0.01}$ \\
 $\Sigma_{00}(*)$           & &$-0.09 \pm 0.06$
 & $\ \ 0.00\pm 0.00$& $\ \ 0.00\pm 0.00$\\
 $\Sigma_{\|\|}(*)$         & &$\ \ 0.10 \pm 0.06$
 & $\ \ 0.01^{+0.00}_{-0.01}$& $-0.00\pm 0.00$\\
 $\Sigma_{\perp\perp}(*)$   & &$-0.01 \pm 0.06$
 & $-0.01\pm 0.01$ & $-0.00\pm 0.00$\\
 $\Sigma_{\| 0}(*)$         & &$-0.11 \pm 0.14$
 & $\ \ 0.06\pm 0.06$ & $\ \ 0.05\pm 0.07$\\
 \hline
 BR(${\overline B}^0 \rightarrow \phi {\overline K}^{*0}$)
                            & &
 & $\ \ (9.0^{+0.5}_{-0.6}\pm1.9)\times 10^{-6}$ & $\ \ (8.7^{+0.5}_{-0.6}\pm 1.8)\times 10^{-6}$\\
 BR(${\overline B}^0 \rightarrow \phi {\overline K}^{*0})_0$&
 & & $\ \ (4.7\pm 0.1\pm 1.0)\times 10^{-6}$& $\ \ (4.5\pm 0.1\pm 0.9)\times 10^{-6}$\\
 BR(${\overline B}^0 \rightarrow \phi {\overline K}^{*0})_\|$
                             & &
 & $\ \ (1.9\pm 0.3\pm 0.4) \times 10^{-6}$& $\ \ (2.2\pm 0.3\pm 0.5) \times 10^{-6}$\\
 BR(${\overline B}^0 \rightarrow \phi {\overline K}^{*0})_\perp$
                             & &
 & $\ \ (2.4\pm 0.3\pm 0.5) \times 10^{-6}$& $\ \ (2.0\pm 0.3\pm 0.5) \times 10^{-6}$\\\
 $A_{CP}({\overline B} \rightarrow \phi {\overline K^*})$
 &$-0.01\pm 0.09$& &$-0.01\pm 0.01$ &$-0.01\pm 0.02$ \\
 $A_{CP}^0=- \Sigma_{00}/\Lambda_{00}$
 &$-0.06\pm 0.10$& &$-0.02\pm 0.02$&$-0.02\pm 0.02$\\
$A_{CP}^\|=- \Sigma_{\| \|}/\Lambda_{\| \|}$
                            & &
                            & $-0.05\pm 0.05$&$-0.01\pm 0.02$\\
$A_{CP}^\perp=- \Sigma_{\perp \perp}/\Lambda_{\perp \perp}$
                            &$-0.10\pm 0.25$&
                          &$\ \ 0.03 \pm 0.03$ &$-0.01\pm 0.01$\\
$A_{CP}^T=- {\Sigma_{\| \|}+\Sigma_{\perp \perp} \over \Lambda_{\|
\|}+\Lambda_{\perp \perp}}$
                           & &
                           & $-0.01\pm 0.01$ &$-0.01\pm 0.02$
\end{tabular}
\end{ruledtabular}
\end{table}
} ${\tilde a}_{23}$ produces sizable contributions to the
transverse amplitudes. $\Lambda_{\|\|} + \Lambda_{\perp\perp} \sim
\L_{00}$ can be understood by following the analysis given in the
first scenario. In this scenario, because the two terms in both
amplitudes ${\overline A}_\|$ and ${\overline A}_\perp$ in
Eq.~(\ref{eqn:hpm1}) contribute constructively, we find
$\Lambda_{\|\|} / \Lambda_{\perp \perp}\approx (\overline
A_{\|}^{NP}/ \overline A_{\perp}^{NP})^2=1.1$.

As for $\phi_{23}=0.14\pm 0.09\ [= (8\pm 5)^\circ]$, we obtain
$A_T^0= 0.27\pm 0.04, \overline A_T^0 =0.22\pm 0.04$, and
accordingly $\Sigma_{\perp 0}= -0.49\pm0.07$, $\Lambda_{\perp
0}=-0.05\pm0.06$. Since the numerical analysis gives the
triple-products $A_T^\|, {\overline A}_T^\| \simeq 0.00\sim 0.01$,
we therefore obtain $\Sigma_{\perp \|}= - \overline A_T^\| -
A_T^\|\simeq -0.01$  and $\Lambda_{\perp \|}=\overline A_T^\|-
A_T^\|\simeq 0$. Note that $\Lambda_{\perp 0,\perp \|}$ are
CP-violating observables. We get $\Lambda_{\| 0}=-0.49\pm 0.07$,
while the SM result is $\Lambda_{\| 0}\simeq 0.30$. For NP related
observables, we obtain $\Sigma_{\| 0} =-\Lambda_{\perp 0} =0.05
\pm 0.07$ but $\Lambda_{\perp\|}\approx
\Sigma_{\lambda\lambda}\approx 0$ which are rather small. Larger
magnitudes of $\L_{\perp 0}$ and $\Sigma_{\| 0}$ are implied for a
larger $|\phi_{23}|$. The BABAR results, displaying
$\arg(\overline A_\perp - A_\perp)\neq 0$ and $\arg(\overline A_\|
- A_\|)\neq 0$, may hint at the existence of the NP weak phase;
consequently, if taking alone the BABAR data, the numerical
analysis yields $\phi_{23}=0.23^{+0.15}_{-0.12}$ such that $A_T^0
(=0.29\pm0.04) \neq \overline A_T^0 (=0.14^{+0.10}_{-0.07}$),
which can be rewritten as $\Sigma_{\perp 0} (=-\overline A_T^0-
A_T^0) = -0.43^{+0.08}_{-0.11}$ and $\Lambda_{\perp 0} (=\overline
A_T^0- A_T^0)=-0.16^{+0.12}_{-0.09}$, and $\Sigma_{\| 0}
(=Re(A_{\|}A_0^*-\overline A_{\|} \overline A_0^*))= 0.15\pm
0.09$. Finally, we get BR($B^0 \rightarrow \phi K^{*0} ) \simeq
(1.22\pm 0.24) \times 10^{-6}$ which is in good agreement with the
world average $(9.5\pm 0.9)\times 10^{-6}$~\cite{hfag}.

\section{Summary and Conclusion} \label{sec:section5}
The large transverse polarization anomaly in the $\overline B \to
\phi \overline K^*$ decays has been observed by BABAR and BELLE.
We resort to the new physics for seeking the possible resolutions.
We have analyzed all possible new-physics four-quark operators.
Following the analysis for the helicities of quarks arising from
various four-quark operators in the $B$ decays, we have found that
there are two classes of operators which could offer resolutions
to the $\overline B\to \phi \overline K^*$ polarization anomaly.
The first class is made of $O_{17,18}$ and $O_{25,26}$ operators
with structures $(1-\gamma_5)\otimes (1-\gamma_5)$ and
$\sigma(1-\gamma_5)\otimes \sigma(1-\gamma_5)$, respectively.
These operators contribute to different helicity amplitudes as
${\overline H_{00}}:{\overline H_{--}}:{\overline H_{++}} \sim
{\cal O}(1/m_b):{\cal O}(1/m_b^2):{\cal O}(1)$. The second class
consists of $O_{15,16}$ and $O_{23,24}$ operators with structures
$(1+\gamma_5)\otimes (1+\gamma_5)$  and $\sigma(1+\gamma_5)\otimes
\sigma(1+\gamma_5)$, respectively, and the resulting amplitudes
are given as ${\overline H_{00}}:{\overline H_{++}}:{\overline
H_{--}} \sim {\cal O}(1/m_b):{\cal O}(1/m_b^2):{\cal O}(1)$.
Moreover, we have shown in Eq.~(\ref{eq:fiez}) that by Fierz
transformation $O_{17,18}$ can be rewritten in terms of
$O_{25,26}$, and $O_{15,16}$ in terms of $O_{23,24}$. For each
class of new physics, we have found that all new physics effects
can be lumped into a sole parameter: $\tilde a_{25}$ (or $\tilde
a_{23}$) in the first (or second) class. Our conclusions are as
follows:
\begin{enumerate}
 \item Two possible experimental
results of polarization phases,
$\arg(A_\perp)-\arg(A_\|)\approx\pi$ or $0$, originating from the
phase ambiguity in data, could be separately accounted for by our
two new-physics scenarios with the presence of a large(r) strong
phase, $\delta_{25}$ (or $\delta_{23})$, and a small weak phase,
$\phi_{25}$ (or $\phi_{23})$. In the fist scenario only the
effective coefficient $\tilde a_{25}$ is relevant, which is
related to $O_{17,18,25,26}$ operators such that ${\overline
H}_{++} \gg {\overline H}_{--}$, while in the second scenario only
the effective coefficient $\tilde a_{23}$ is relevant, which is
associated with $O_{15,16,23,24}$ operators such that ${\overline
H}_{--} \gg {\overline H}_{++}$.  Note that if simultaneously
considering the six parameters $|\tilde a_{25}|, \delta_{25},
\phi_{25},|\tilde a_{23}|, \delta_{23}, \phi_{23}$ in the fit, the
final results still converge to the above two scenarios.
 \item We obtain $\L_{\|\|} \simeq 0.8 \L_{\perp\perp}$
in the first scenario, but $\L_{\|\|} \gtrsim \L_{\perp\perp}$ in
the second scenario.
 \item Our numerical analysis yields
$\overline A_T^{\|}, A_T^{\|} \approx 0.10$ and
$\Sigma_{\perp\|}\approx -0.19$ in the first scenario, but gives
$\overline A_T^{\|}, A_T^{\|}\simeq 0.01$ and
$\Sigma_{\perp\|}\simeq -0.01$ in the second scenario. These two
scenarios can thus be distinguished. Furthermore, a larger
magnitude of the weak phase, $\phi_{25}$ or $\phi_{23}$, can
result in sizable $\Lambda_{\perp 0}, \Sigma_{\| 0}$. As displayed
in Table~\ref{tab:observables}, we obtain $\Lambda_{\perp 0}\simeq
- \Sigma_{\| 0}\simeq -0.05$ for $\phi_{25,(23)}=-0.11\ (0.14)$.
 \item The NP related observations
$\Sigma_{00, \|\|, \perp\perp}, \Lambda_{\perp\|}$ are only
marginally affected by weak phases $\phi_{25,23}$.
 \item We obtain BR($B \rightarrow \phi K^{*})\simeq (1.3\pm 0.3)\times
10^{-6}$ in two scenarios. Note that we have used the rescaled
LCSR form factors in Ref.~\cite{alisafir,Ball:1998kk}, where
smaller values for form factors were used in explaining $B
\rightarrow K^* \gamma, X_s \gamma$ data~\cite{Ali:2004hn}.

\end{enumerate}

\begin{acknowledgments}

 We are grateful to Hai-Yang Cheng and Kai-Feng Chen for useful
discussions. We thank Andrei Gritsan and Alex Kagan for
many helpful comments on the manuscript.
This work was supported in part by the National
Science Council of R.O.C. under Grant Nos: NSC92-2112-M-033-014,
NSC93-2112-M-033-004, and NSC93-2811-M-033-004.

\end{acknowledgments}

\vskip2cm
\appendix
\section{}\label{app:a}
The LCDAs of the vector meson relevant for the present study are
given by~\cite{Beneke:2000wa}
\begin{eqnarray}
  &&\langle V(P',\epsilon)|\bar q_1(y) \gamma_\mu q_2(x)|0\rangle
  = f_V m_V \, \int_0^1
      du \,  e^{i (u \, p'\cdot y +
    \bar u p' \cdot x)}
   \left\{p^\prime_\mu \,
    \frac{\epsilon^*\cdot z}{p'\cdot z} \, \Phi_\parallel(u)
         +\epsilon_\perp^*{}_\mu \, g_\perp^{(v)}(u)
         \right\}, \nonumber\\ \label{evendef1} \\
  &&\langle V(P',\epsilon)|\bar q_1(y) \gamma_\mu\gamma_5
  q_2(x)|0\rangle \nonumber\\
  && \ \ \
  = -  f_V \Bigg(1-{f_V^{T} \over f_V}{m_{q_1}+m_{q_2}\over m_V}\Bigg)
  m_V \,\epsilon_{\mu\nu\rho\sigma} \,
      \epsilon^*{}^\nu p^{\prime\rho} z^\sigma \,
    \int_0^1 du \,  e^{i (u \, p'\cdot y +
    \bar u p' \cdot x)} \,
       \frac{g_\perp^{(a)}(u)}{4}, \label{evendef2} \\
  &&\langle V(P',\epsilon)|\bar q_1(y) \sigma_{\mu\nu} q_2(x)
            |0\rangle
  =- i f_V^{T} \,\int_0^1 du \, e^{i (u \, p'\cdot y +
    \bar u p' \cdot x)} \,
(\epsilon^*_\perp{}_\mu p_{\nu}^\prime -
  \epsilon_\perp^*{}_\nu  p^\prime_{\mu}) \, \Phi_\perp(u),\label{evendef3}
\end{eqnarray}
where $z=y-x$ with $z^2=0$, and we have introduced the light-like vector
$p'_\mu=P'_\mu-m_V^2 z_\mu/(2 P'\cdot z)$ with the meson's
momentum ${P'}^2=m_V^2$. Here the
longitudinal and transverse {\it projections} of the polarization
vectors are defined as
 \bea && \epsilon^*_\parallel{}_\mu \equiv
     \frac{\epsilon^* \cdot z}{P'\cdot z} \left(
      P'_\mu-\frac{m_V^2}{P'\cdot z} \,z_\mu\right), \qquad
 \epsilon^*_\perp{}_\mu
        = \epsilon^*_\mu -\epsilon^*_\parallel{}_\mu .
\eea
 Note that these are not exactly the polarization vectors of the
 vector meson.
 In the QCDF calculation, the LCDAs of the meson appear in the following way
  \bea
  &&  \langle V(P',\epsilon)|\bar q_{1\,\alpha}(y) \, q_{2\, \delta}(x)|0\rangle
= \frac{1}{4} \, \int_0^1 du \,  e^{i (u \, p'\cdot y +
    \bar u p' \cdot x)}
\nonumber\\[0.1cm]
  && \quad \times\,\Bigg\{ f_V m_V \left(
    p^\prime_\mu \, \frac{\epsilon^*\cdot z}{p'\cdot z} \,
    \Phi_\parallel(u) +  \not\! \epsilon^*_\perp \,
    g_\perp^{(v)}(u) +  \epsilon_{\mu\nu\rho\sigma} \,
    \epsilon^*{}^\mu  p^{\prime\rho} z^\sigma \, \gamma^\mu\gamma_5
    \, \frac{g_\perp^{(a)}(u)}{4}\right)
\nonumber \\[0.1em]
  && \qquad \,\,\,
  + \,f^{T}_V \not\!\epsilon_\perp^* \not\! p^\prime \,
  \Phi_\perp(u)\Bigg\}_{\delta\alpha}\!\!.
 \eea
  Note that to perform the calculation in the momentum space, we
  first represent the above equation in terms of $z$-independent
 variables, $P'$ and $\epsilon^*$. Then, the light-cone
 projection operator of a light vector meson in the momentum space
 reads
\begin{equation}
  M_{\delta\alpha}^V =  M_{\delta\alpha}^V{}_\parallel +
   M_{\delta\alpha}^V{}_\perp\,,
\label{rhomeson2}
\end{equation}
with the longitudinal projector
 \bea M^V_\parallel &=& \frac{f_V}{4} \,
\frac{m_V(\epsilon^*\cdot
  n_+)}{2}
 \not\! n_- \,\Phi_\parallel(u) \Bigg|_{k=up'}\,,
\eea
 and the transverse projector
 \bea
M^V_\perp &=& \frac{f^{T}_V}{4} \,\not\! \epsilon^*_\perp \not\!
p^\prime \,
   \Phi_\perp(u)\nonumber\\
&&  +\frac{f_Vm_V}{4} \,\Bigg\{\not\! \epsilon^*_\perp\,
g_\perp^{(v)}(u) -  \, \int_0^u dv\, (\Phi_\parallel(v) -
g_\perp^{(v)}(v))  \
       \not\! p^\prime \, \epsilon^*_{\perp\mu} \,\frac{\partial}{\partial
         k_{\perp\mu}}
\cr && + \,i \epsilon_{\mu\nu\rho\sigma} \,
        \epsilon_\perp^{*\nu} \, \gamma^\mu\gamma_5
         \left[n_-^\rho n_+^\sigma \,{1\over 8}\frac{dg_\perp^{(a)}(u)}{du}-
          p^{\prime\rho}\,\frac{g_\perp^{(a)}(u)}{4} \, \frac{\partial}{\partial
         k_\perp{}_\sigma}\right]
 \Bigg\}
 \, \Bigg|_{k=up'},
\eea
 where $n_-^\mu\equiv (1,0,0,-1), n_+^\mu\equiv (1,0,0,1)$,
$k_\perp$ is the transverse momentum of the $q_1$ quark in the
vector meson, and the polarization vectors of the vector meson are
\begin{equation}
\epsilon_\perp^\mu \equiv \epsilon^\mu -
\frac{\epsilon\cdot n_+}{2}\,n_-^\mu- \frac{\epsilon\cdot
n_-}{2}\,n_+^\mu .
\end{equation}
In the present study, we only consider the leading contribution in
$\Lambda_{\rm QCD}/m_b$ for $M_{\|}^V$. In Eqs.~(\ref{evendef1}),
(\ref{evendef2}) and (\ref{evendef3}), $\Phi_{\|},\Phi_{\perp}$
are twist-2 LCDAs, while $g_{\perp}^{(v)}, g_{\perp}^{(a)}$ are
twist-3 ones. Applying the equation of motions to LCDAs, one can
obtain the following Wandzura-Wilczek relations
 \bea
  g_\perp^{(v)}(u) &=& \frac12 \left[ \,\int_0^u
    \frac{\Phi_\parallel(v)}{\bar v} \, dv  + \int_u^1
    \frac{\Phi_\parallel(v)}{v}\, dv  \right] +\ldots \,,\label{ww1}
 \\
  g_\perp^{(a)}(u) &=& 2 \left[ \bar u \int_0^u
    \frac{\Phi_\parallel(v)}{\bar v}\, dv  + u \, \int_u^1
    \frac{\Phi_\parallel(v)}{v}\, dv  \right] +\ldots \,,
\label{ww2} \eea
 where the ellipses in Eqs.~(\ref{ww1}) and (\ref{ww2})
  denote additional contributions from
three-particle distribution amplitudes containing gluons and terms
proportional to light quark masses, which we do not consider here.
Eqs.~(\ref{ww1}) and (\ref{ww2}) further give
 \bea
{1\over 4}\frac{d g_\perp^{(a)}(u)}{du} + g_\perp^{(v)}(u) &=& \int_u^1
 \frac{\Phi_\parallel(v)}{v} \, dv +\ldots\,,
\label{ww3}\\
 \int_0^u \, (\Phi_\parallel(v)-g_\perp^{(v)}(v)) \, dv &=&
   \frac12 \left[ \bar u \int_0^u
    \frac{\Phi_\parallel(v)}{\bar v}\, dv  - u \, \int_u^1
    \frac{\Phi_\parallel(v)}{v}\, dv  \right] +\ldots\,,
\label{ww4} \eea
 After considering Eqs.~(\ref{ww1}), (\ref{ww2}), (\ref{ww3})
 and (\ref{ww4}), $G_g^{\pm}$ in
Eq.~(\ref{eq:cg}) are actually equal to zero.

\end{document}